\documentclass[onecolumn,11pt,draftcls]{IEEEtran}

\usepackage[noadjust]{cite}
\usepackage{amsmath}
\usepackage{amsfonts}
\usepackage{graphicx}
\usepackage{float}
\usepackage{subfigure}
\usepackage{epsfig}
\usepackage{amsmath}
\usepackage{theorem}
\usepackage{amsmath}
\usepackage{mathrsfs}
\usepackage{amsfonts}
\usepackage{amssymb}
\usepackage{mathrsfs}
\usepackage{euscript}
\usepackage{multirow}
\usepackage{cite}
\usepackage{url}
\usepackage{color}

\newtheorem{remark}{Remark}
\begin{document}
\title{Maximizing the minimum achievable secrecy rate of two-way relay networks using the null space beamforming method}
\author
{\IEEEauthorblockN{Erfan khordad, Soroush Akhlaghi*, Meysam Mirzaee}
\\ \IEEEauthorblockA{Electrical and Electronic Engineering Department, Shahed University, Tehran, Iran\\
{*Akhlaghi@shahed.ac.ir}}}
\maketitle
\begin{abstract}
This paper concerns maximizing the minimum achievable secrecy rate of a two-way relay network in the presence of an eavesdropper, in which two nodes aim to exchange messages in two hops, using a multi-antenna relay. Throughout the first hop, the two nodes simultaneously transmit their messages to the relay. In the second hop, the relay broadcasts a combination of the received information to the users such that the transmitted signal lies in the null space of the eavesdropper's channel; this is called null space beamforming~(NSBF). The best NSBF matrix for maximizing the minimum achievable secrecy rate is studied, showing that the problem is not convex in general. To address this issue, the problem is divided into three sub-problems: a close-to-optimal solution is derived by  using the semi-definite relaxation~(SDR) technique. Simulation results demonstrate the superiority of the proposed method w.r.t. the most well-known method addressed in the literature.
\end{abstract}
\section{INTRODUCTION}\label{sec:intro}
Secure data transmission is regarded as one of the main challenges in wireless networks. This is due to the nature of the wireless medium, where any illegitimate node can hear over-the-air messages. There have been some attempts to alleviate this problem to protect transmitted information against eavesdropping. In a landmark paper, Wyner introduced the notion of physical layer security and showed that as long as the source-to-eavesdropper channel is a degraded version of the source-to-destination channel, the data can be securely transmitted~\cite{Wyner}. Also, Wyner defined the secrecy capacity as the maximum transmission rate below which the private message could be sent reliably from the source to the legitimate destination.

There have been some attempts to explore the secrecy capacity of a variety of network topologies, including so-called relay-assisted networks which are regarded as a promising solution to enhancing the quality of wireless networks~\cite{ucdpisd,khajenouri}. There have also been some attempts to seek effective coding strategies at relays to enhance physical layer security \cite{treccfs,orsfplsicwn,soodharswrswec}.

In addition, two-way relaying is mostly regarded as a spectrally efficient solution, e.g., see~\cite{sepfhfrc,wncbaaffbdtf}. It is desirable to investigate the reliability of such networks from the physical layer security perspective. Motivated by this, some research has  explored the security of two-way relaying~\cite{dbfplsotwrn,base,hcbajfplsotwrn}. Also, the impact of multi-antenna relaying on physical layer security has been investigated~\cite{rcbaandfplsiamamrn,jsrpapafsaafmrn,cjfscimrn}.

It should be noted that different strategies may be employed at relays; the literature mostly addresses the amplify-and-forward~(AF) strategy. This is largely due to the simplicity of this strategy, since it merely transmits a scaled version of the incoming signal to the destination. Nevertheless, when a relay is an untrusted node of the network, the AF strategy is a promising approach, because the relay is not required to go through the information bits. The secrecy capacity of a two-way channel incorporating some single-antenna AF relays has been investigated in~\cite{base}.

One of the main objectives pursued in~\cite{base} was obtaining the scaling factors associated with the relays to maximize the minimum achievable secrecy rates (ASRs) of transmitting ends, assuming the relays are subject to a sum power constraint. Moreover, the relay-to-eavesdropper channel information is known globally at the relays; thus, the transmitted signal vector from relays resides in the null space of the relay-to-eavesdropper channel. Given this, the eavesdropper does not receive any information throughout the second hop. This enhances the overall secrecy capacity. Accordingly,~\cite{base} demonstrated that the incorporated NSBF simplifies the underlying optimization problem, where the sub-optimal solution is numerically computed.

Referring to~\cite{Aylin}, representing the secrecy rates of transmitting ends by $R_1$ and $R_2$, the secrecy rate region is identified by applying three constraints on $R_1$ and $R_2$ as $R_1\leq R_{1}^{up}$, $R_2\leq R_{2}^{up}$, and $R_1+R_2\leq R_{sum}^{up}$. However, \cite{base} does not take the sum secrecy rate constraint into account; hence, the obtained result will be an upper bound, as the obtained rates may not reside inside the secrecy rate region. This is the main drawback of~\cite{base}.

Moreover, the AF relay can be equipped with multiple antennas to steer the transmitted signal vector along a desirable direction to maintain specific criteria such as keeping the eavesdropper as ignorant as possible of the transmitted signal vector by sending the information vector in the null space of the eavesdropper's channel. The objective of this paper is to investigate the ASR of a two-way multi-antenna relay assisted network. The main contributions of this paper can be summarized as follows.

First, assuming an NSBF matrix is employed at the relay, the best beamforming matrix, in terms of maximizing the minimum ASR, is numerically derived. This maximization is performed in the ASR region obtained in \cite{Aylin} and among all NSBF matrices. It should be noted that due to taking the sum secrecy rate constraint into account, the method proposed in \cite{base} cannot be employed.

To tackle the problem, depending on the channel realizations, and based on the possibility of locating the optimal point on each of three borders of the ASR region, three sub-problems are formulated, where the optimal solution is the best one of the group. To this end, it is shown that these sub-problems are not convex in general. However, they can be reformulated as three semi-definite programming (SDP) problems using the SDR method. They can be solved using well-known software packages such as the CVX package which is employed in the current work~\cite{cvx}.

Although the solution of the equivalent SDR problem is an upper bound of the main problem, a close-to-optimal solution is proposed that is shown to closely follow the upper bound. Numerical results demonstrate the superiority of the proposed method in terms of the minimum ASR as compared with the beamforming vector obtained from~\cite{base}.

The rest of this paper is organized as follows: Section II presents the system model and formulates the individual secrecy rates as well as sum secrecy rate w.r.t. the beamforming matrix at the relay. Section III discusses the optimization problem and presents further steps towards obtaining a close-to-optimal solution. The results are presented in Section IV. Finally, Section V concludes the paper with findings.

Throughout the paper, complex conjugate, transpose, and Hermitian transpose are denoted by $(.)^*$, $(.)^\text{T}$, and $(.)^\text{H}$, respectively. Also $\otimes$, $\text{Tr(.)}$, $\text{vec(.)}$ indicate the kronecker product, the matrix trace operator, and the vectorisation operator. Upper and lower bold face letters are used for matrices and vectors, respectively. The absolute value, the 2-norm value, and the $N\times{N}$ identity matrix are represented by $\left|.\right|$, $\|.\|$, and $\textbf{I}$, respectively. Finally, ${{\left[ \,x\, \right]}^{+}}$ means $\max \left\{ 0\,\,,\,\,x \right\}$, $I\left(a\,;b\right)$ denotes the mutual information of two random variables $a$ and $b$, and $diag(a,b)$ indicates a diagonal matrix with diagonal elements $a$ and $b$. 
\section{SYSTEM MODEL}\label{sec:sysmod}
\begin{figure}[t]
\centering
 \includegraphics[scale=1]{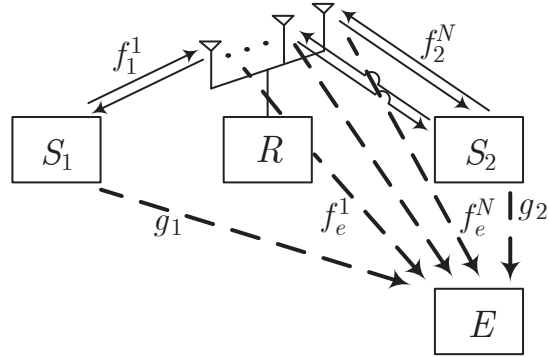}
  \caption{The structure of the considered network}\label{fig_model}
\end{figure}

We consider a wireless two-way relay network consisting of two single-antenna nodes $S_1$ and $S_2$, an AF $N$-antenna relay $R$, and a single-antenna eavesdropper node $E$ as depicted in Fig. \ref{fig_model}. We assume that the eavesdropper is an active and untrusted node of the network, thus, in addition to nodes $S_1$ and $S_2$, the channel state information (CSI) of $E$ is also available at the relay. This is a common assumption in similar research including~\cite{dbfplsotwrn,base}.

It is assumed that $S_1$ and $S_2$ aim at exchanging messages through the use of a relay in the middle of transmission in a two-hop fashion, where throughout the first hop, both nodes $S_1$ and $S_2$ transmit signals $x_1$ and $x_2$, respectively, with powers $\text{P}_1$ and $\text{P}_2$ to the relay. Thus, the signal received at the relay is $\mathbf{r}={{\mathbf{f}}_{1}}{{x}_{1}}+{{\mathbf{f}}_{2}}{{x}_{2}}+{\mathbf{n}_{R}}$, and the following signal is received at the eavesdropper,
\begin{equation}\label{ye1}
y_{{E}}^{\left( 1 \right)}={{g}_{1}}{{x}_{1}}+{{g}_{2}}{{x}_{2}}+n_{{E}}^{\left( 1 \right)},
\end{equation}
where ${{\mathbf{f}}_{1}}={{\left[ f_{1}^{1},\,f_{1}^{2}\,,\,.\,.\,.\,,\,f_{1}^{N} \right]}^{\text{T}}}$ and ${{\mathbf{f}}_{2}}={{\left[ f_{2}^{1},\,f_{2}^{2}\,,\,.\,.\,.\,,\,f_{2}^{N} \right]}^{\text{T}}}$ are reciprocal quasi-static channel gain vectors associated with $S_1-R$ and $S_2-R$ channels. Also, $g_1$ and $g_2$ denote, respectively, the channel gains from $S_1$ and $S_2$ to the eavesdropper. Moreover, ${{\mathbf{n}}_{{R}}}~\sim ~\mathcal{CN}(\textbf{0},\sigma_R^{2}{{\mathbf{I}}})$ and $n_{{{E}}}^{\left( 1 \right)}~\sim ~\mathcal{CN}(0,\sigma _{{{E}},1}^{2})$ are received noises at the relay and the eavesdropper, respectively.

Then, during the second hop, the relay applies beamforming matrix $\mathbf{W}$ to the received signal vector $\mathbf{r}$ and transmits resulting vector $\mathbf{s}=\mathbf{Wr}$ to the nodes $S_1$ and $S_2$. As a result, the received signals at $S_1$ and $S_2$ can be expressed as,
\begin{equation}\label{yr}
{{y}^{r}_{\text{i}}}={\mathbf{f}_{\text{i}}^\text{T}\mathbf{W}{{\mathbf{f}}_{\text{i}}}{{x}_{\text{i}}}}+\mathbf{f}_{\text{i}}^\text{T}\mathbf{W}{{\mathbf{f}}_{3-\text{i}}}{{x}_{3-\text{i}}}+\mathbf{f}_{\text{i}}^\text{T}\mathbf{W}{{\mathbf{n}}_{{R}}}+{{n}_{\text{i}}},
\end{equation}
where ${y}^{r}_{\text{i}}$ for $\text{i}=1,2$ represents the received signal at node $S_\text{i}$. Also,
${{n}_{\text{i}}}\sim\mathcal{CN}(0,\sigma _{\text{i}}^{2})$ is the received noise at the reception side of $S_\text{i}$ with zero mean and variance $\sigma _{\text{i}}^{2}$. Referring to (\ref{yr}), it should be noted that the transmission ends should have access to the relay's beamforming matrix to subtract the self-interference term, i.e., ${\mathbf{f}_{\text{i}}^\text{T}\mathbf{W}{{\mathbf{f}}_{\text{i}}}{{x}_{\text{i}}}}$. This can be done by sending this matrix at the start of each transmission block. Thus, knowing $\mathbf{f}_{\text{i}}$ and $\bold{W}$ at node $S_\text{i}$, the self-interference term is completely known at this node, hence, it can be subtracted from ${{y}^{r}_{\text{i}}}$, which gives,
\begin{equation}\label{yi}
{y}_{\text{i}}=\mathbf{f}_{\text{i}}^{\text{T}}\mathbf{W}{{\mathbf{f}}_{3-\text{i}}}{{x}_{3-\text{i}}}+\mathbf{f}_{\text{i}}^\text{T}\mathbf{W}{{\mathbf{n}}_{{R}}}+{{n}_{\text{i}}}.
\end{equation}
Also, in the second hop, the received signal at node $E$ becomes,
\begin{equation}\label{yE2}
y_{{{E}}}^{\left( 2 \right)}=\mathbf{f}_{e}^\text{T}\mathbf{W}{{\mathbf{f}}_{1}}{{x}_{1}}+\mathbf{f}_{e}^{\text{T}}\mathbf{W}{{\mathbf{f}}_{2}}{{x}_{2}}+\mathbf{f}_{e}^{\text{T}}\mathbf{W}{{\mathbf{n}}_{{R}}}+n_{{{E}}}^{\left(2\right)},
\end{equation}
where ${{\mathbf{f}}_{{e}}}={{\left[ f_{e}^{1},\,f_{e}^{2}\,,\,.\,.\,.\,,\,f_{e}^{N} \right]}^{\text{T}}}$ contains the relay-eavesdropper channel coefficients and $n_{{{E}}}^{\left( 2 \right)}$ $\sim $ $\mathcal{CN}(0,\sigma _{E,2}^{2})$ is the received noise at node $E$ .

Noting~(\ref{ye1}) and (\ref{yE2}), a useful expression of the received signals at $E$ during the two hops can be written as,
\begin{equation}\label{YE}
\begin{aligned}
{{\mathbf{y}}_{{E}}}=\left[ \begin{matrix}
   y_{E}^{(1)}  \\
   y_{E}^{(2)}  \\
\end{matrix} \right]
=\underbrace{\left[
\begin{matrix}
{{g}_{1}} & {{g}_{2}}  \\
\mathbf{f}_{e}^{\text{T}}\mathbf{W}{{\mathbf{f}}_{1}} & \mathbf{f}_{e}^{\text{T}}\mathbf{W}{{\mathbf{f}}_{2}}  \\
\end{matrix} \right]}_{\mathbf{U}}{\left[ \begin{matrix}
   {{x}_{1}}  \\
   {{x}_{2}}  \\
\end{matrix} \right]}+{\left[ \begin{matrix}
n_{{{E}}}^{\left( 1 \right)}  \\
\mathbf{f}_{e}^{\text{T}}\mathbf{W}{{\mathbf{n}}_{{R}}}+n_{{{E}}}^{\left( 2 \right)}\\
\end{matrix} \right]},
\end{aligned}
\end{equation}

Considering the ASR received at node $S_\text{i}$ is denoted by $R_\text{i}$, in \cite{Aylin}, under Gaussian codebook assumption and stochastic encoders, it is shown that the ASR region of a two-way wiretap channel is identified as follows,
\begin{align}\label{ASR Region}
  \mathcal{R} = &\Biggl\{ (R_1,R_2):  {{R}_{1}}\le  \frac{1}{2}{{\left[ I\left( {{x}_{2}};{{y}_{1}} \right)-I\left( {{x}_{2}};{{\mathbf{y}}_{{E}}} \right) \right]}^{+}} \nonumber,
{{R}_{2}}\le \frac{1}{2}{{\left[ I\left( {{x}_{1}};{{y}_{2}} \right)-I\left( {{x}_{1}};{{\mathbf{y}}_{{E}}} \right) \right]}^{+}} \nonumber \\
\hspace{-1cm}{{R}_{1}}+{{R}_{2}}&\le \frac{1}{2}{{\left[ I\left( {{{x}}_{1}};{{{y}}_{2}} \right)+I\left( {{x}_{2}};{{y}_{1}} \right)-I\left( {{{x}}_{1}}{,}{{{x}}_{2}};{{\mathbf{y}}_{{E}}} \right) \right]}^{+}}\Biggr\},
\end{align}
where the half factor is due to the fact that the transmission is carried out in two hops. To compute the mutual information on the right side of the above constraints, using (\ref{yi}) and (\ref{YE}), it follows,
\begin{equation}\label{Ix2y1}
I\left({{x}_{2}};{{y}_{1}}\right)={{\log}_{2}}\left(1+\frac{{{\text{P}}_{2}}{{\left|\mathbf{f}_{1}^{\text{\text{T}}}\mathbf{W}{{\mathbf{f}}_{2}}\right|}^{2}}}{\sigma_{{R}}^{2}{{\|\mathbf{f}_{1}^{\text{\text{T}}}\mathbf{W}\|}^{2}}+\sigma_{1}^{2}} \right),
\end{equation}
\begin{equation}\label{Ix1y2}
I\left({{x}_{1}};{{y}_{2}}\right)={{\log}_{2}}\left(1+\frac{{{\text{P}}_{1}}{{\left|\mathbf{f}_{2}^{\text{T}}\mathbf{W}{{\mathbf{f}}_{1}}\right|}^{2}}}{\sigma_{{R}}^{2}{{\|\mathbf{f}_{2}^{\text{T}}\mathbf{W}\|}^{2}}+\sigma_{2}^{2}} \right),
\end{equation}
\begin{equation}\label{Ix1x2YE}
{I}\left( {{{x}}_{1}},{{{x}}_{2}};{{\mathbf{y}}_{{E}}} \right)={{\log }_{2}}\det \left({{\mathbf{I}}_{2}}+\frac{{{\mathbf{U}}}{{\mathbf{M}}}\mathbf{U}^{\text{H}}}{{{\mathbf{L}}}}\right),
\end{equation}
where (\ref{Ix1x2YE}) comes from an equivalent MIMO system model. Also, $\mathbf{I}_2$ indicates the $2\times2$ identity matrix, ${\mathbf{M}}= diag(\text{P}_1,\text{P}_2)$, and ${\mathbf{L}}= diag(\sigma _{{{E}},1}^{2},\sigma _{{{E}},2}^{2}+\sigma _{{R}}^{2}{{\| \mathbf{f}_{e}^{\text{T}}\mathbf{W} \|}^{2}})$. Furthermore, referring to (\ref{YE}), and extracting the mutual information using the entropy function, one can arrive at equations (\ref{botpage1}) and (\ref{botpage2}).
\begin{figure*}[t]
\normalsize
\begin{equation}\label{botpage1}
\begin{aligned}
I\left( {{x}_{2}};{{\mathbf{y}}_{{E}}} \right)=
{{\log}_{2}}&\Bigg[{1+\frac{{{\left|{{g}_{2}}\right|}^{2}}{{\text{P}}_{2}}\sigma_{{R}}^{2}{{\left\|\mathbf{f}_{e}^{\text{T}}\mathbf{W}\right\|}^{2}}+{{\left|{{g}_{2}}\right|}^{2}}{{\text{P}}_{2}}{{\left|\mathbf{f}_{e}^{\text{T}}\mathbf{W}{{\mathbf{f}}_{1}}\right|}^{2}}{{\text{P}}_{1}}+{{\left|{{g}_{2}}\right|}^{2}}{{\text{P}}_{2}}\sigma_{{E,2}}^{2}+{{\left|{{g}_{1}}\right|}^{2}}{{\text{P}}_{1}}{{\left|\mathbf{f}_{e}^{\text{T}}\mathbf{W}{{\mathbf{f}}_{2}}\right|}^{2}}{{\text{P}}_{2}}}{{{\left|{{g}_{1}}\right|}^{2}}{{\text{P}}_{1}}\sigma_{{R}}^{2}{{\left\|\mathbf{f}_{e}^{\text{T}}\mathbf{W}\right\|}^{2}}+{{\left|{{g}_{1}}\right|}^{2}}{{\text{P}}_{1}}\sigma_{{E,2}}^{2}+\sigma_{{E,1}}^{2}{{\left|\mathbf{f}_{e}^{\text{T}}\mathbf{W}{{\mathbf{f}}_{1}}\right|}^{2}}{{\text{P}}_{1}}\,+\sigma_{{E,1}}^{2}\sigma_{{R}}^{2}{{\left\|\mathbf{f}_{e}^{\text{T}}\mathbf{W}\right\|}^{2}}+\sigma_{{E,1}}^{2}\sigma_{{E,2}}^{2}}}\\
&+\frac{\sigma_{{E,1}}^{2}{{\left|\mathbf{f}_{e}^{\text{T}}\mathbf{W}{{\mathbf{f}}_{2}}\right|}^{2}}{{\text{P}}_{2}}-{{g}_{1}}{{\left(\mathbf{f}_{e}^{\text{T}}\mathbf{W}{{\mathbf{f}}_{1}}\right)}^{*}}{{\text{P}}_{1}}g_{2}^{*}\left(\mathbf{f}_{e}^{\text{T}}\mathbf{W}{{\mathbf{f}}_{2}}\right){{\text{P}}_{2}}-g_{1}^{*}\left(\mathbf{f}_{e}^{\text{T}}\mathbf{W}{{\mathbf{f}}_{1}}\right){{\text{P}}_{1}}{{g}_{2}}{{\left(\mathbf{f}_{e}^{\text{T}}\mathbf{W}{{\mathbf{f}}_{2}}\right)}^{*}}{{\text{P}}_{2}}}{{{\left|{{g}_{1}}\right|}^{2}}{{\text{P}}_{1}}\sigma_{{R}}^{2}{{\left\|\mathbf{f}_{e}^{\text{T}}\mathbf{W}\right\|}^{2}}+{{\left|{{g}_{1}}\right|}^{2}}{{\text{P}}_{1}}\sigma_{{E,2}}^{2}+\sigma_{{E,1}}^{2}{{\left|\mathbf{f}_{e}^{\text{T}}\mathbf{W}{{\mathbf{f}}_{1}}\right|}^{2}}{{\text{P}}_{1}}\,+\sigma_{{E,1}}^{2}\sigma_{{R}}^{2}{{\left\|\mathbf{f}_{e}^{\text{T}}\mathbf{W}\right\|}^{2}}+\sigma_{{E,1}}^{2}\sigma_{{E,2}}^{2}}\Bigg]
\end{aligned}
\end{equation}
\end{figure*}
\begin{figure*}[t]
\normalsize
\begin{equation}\label{botpage2}
\begin{aligned}
I\left( {{x}_{1}};{{\mathbf{y}}_{{E}}} \right)=
{{\log}_{2}}&\Bigg[{1+\frac{{{\left|{{g}_{1}}\right|}^{2}}{{\text{P}}_{1}}\sigma_{{R}}^{2}{{\left\|\mathbf{f}_{e}^{\text{T}}\mathbf{W}\right\|}^{2}}+{{\left|{{g}_{1}}\right|}^{2}}{{\text{P}}_{1}}{{\left|\mathbf{f}_{e}^{\text{T}}\mathbf{W}{{\mathbf{f}}_{2}}\right|}^{2}}{{\text{P}}_{2}}+{{\left|{{g}_{1}}\right|}^{2}}{{\text{P}}_{1}}\sigma_{E,2}^{2}+{{\left|{{g}_{2}}\right|}^{2}}{{\text{P}}_{2}}{{\left|\mathbf{f}_{e}^{\text{T}}\mathbf{W}{{\mathbf{f}}_{1}}\right|}^{2}}{{\text{P}}_{1}}}{{{\left|{{g}_{2}}\right|}^{2}}{{\text{P}}_{2}}\sigma_{{R}}^{2}{{\left\|\mathbf{f}_{e}^{\text{T}}\mathbf{W}\right\|}^{2}}+{{\left|{{g}_{2}}\right|}^{2}}{{\text{P}}_{2}}\sigma_{{E,2}}^{2}+\sigma_{{E,1}}^{2}{{\left|\mathbf{f}_{e}^{\text{T}}\mathbf{W}{{\mathbf{f}}_{2}}\right|}^{2}}{{\text{P}}_{2}}\,+\sigma_{{E,1}}^{2}\sigma_{{R}}^{2}{{\left\|\mathbf{f}_{e}^{\text{T}}\mathbf{W}\right\|}^{2}}+\sigma_{{E,1}}^{2}\sigma_{{E,2}}^{2}}}\\
&+\frac{\sigma_{{E,1}}^{2}{{\left|\mathbf{f}_{e}^{\text{T}}\mathbf{W}{{\mathbf{f}}_{1}}\right|}^{2}}{{\text{P}}_{1}}-{{g}_{1}}{{\left(\mathbf{f}_{e}^{\text{T}}\mathbf{W}{{\mathbf{f}}_{1}}\right)}^{*}}{{\text{P}}_{1}}g_{2}^{*}\left(\mathbf{f}_{e}^{\text{T}}\mathbf{W}{{\mathbf{f}}_{2}}\right){{\text{P}}_{2}}-g_{1}^{*}\left(\mathbf{f}_{e}^{\text{T}}\mathbf{W}{{\mathbf{f}}_{1}}\right){{\text{P}}_{1}}{{g}_{2}}{{\left(\mathbf{f}_{e}^{\text{T}}\mathbf{W}{{\mathbf{f}}_{2}}\right)}^{*}}{{\text{P}}_{2}}}{{{\left|{{g}_{2}}\right|}^{2}}{{\text{P}}_{2}}\sigma_{{R}}^{2}{{\left\|\mathbf{f}_{e}^{\text{T}}\mathbf{W}\right\|}^{2}}+{{\left|{{g}_{2}}\right|}^{2}}{{\text{P}}_{2}}\sigma_{{E,2}}^{2}+\sigma_{{E,1}}^{2}{{\left|\mathbf{f}_{e}^{\text{T}}\mathbf{W}{{\mathbf{f}}_{2}}\right|}^{2}}{{\text{P}}_{2}}\,+\sigma_{{E,1}}^{2}\sigma_{{R}}^{2}{{\left\|\mathbf{f}_{e}^{\text{T}}\mathbf{W}\right\|}^{2}}+\sigma_{{E,1}}^{2}\sigma_{{E,2}}^{2}}\Bigg]
\end{aligned}
\end{equation}
\hrulefill
\end{figure*}

Moreover, defining $\mathbf{w} \triangleq \text{vec}(\textbf{W}^\text{H})$, it follows,
\begin{equation}\label{f1Twf2power2}
\begin{aligned}
{{\left|\mathbf{f}_{1}^{\text{T}}\mathbf{W}{{\mathbf{f}}_{2}}\right|}^{2}}&=\,\text{Tr}\left(\mathbf{f}_{1}^{\text{T}}\mathbf{W}{{\mathbf{f}}_{2}}\mathbf{f}_{2}^{\text{H}}{{\mathbf{W}}^{\text{H}}}\mathbf{f}_{1}^{*}\right) \overset{\left(\text{a}\right)}{\mathop{=}}\,\text{Tr}\left(\mathbf{W}{{\mathbf{f}}_{2}}\mathbf{f}_{2}^{\text{H}}{{\mathbf{W}}^{\text{H}}}\mathbf{f}_{1}^{*}\mathbf{f}_{1}^{\text{T}}\right)\overset{\left(\text{b}\right)}{\mathop{=}}\,\,\,\text{ve}{{\text{c}}^{\text{H}}}\left({{\mathbf{W}}^{\text{H}}}\right)\text{vec}\left({{\mathbf{f}}_{2}}\mathbf{f}_{2}^{\text{H}}{{\mathbf{W}}^{\text{H}}}\,\mathbf{f}_{1}^{*}\mathbf{f}_{1}^{\text{T}}\right)\\
&\overset{\left(\text{c}\right)}{\mathop{=}}\,\,\,{{\mathbf{w}}^{\text{H}}}\left(\mathbf{f}_{1}^{{}}\mathbf{f}_{1}^{\text{H}}\otimes\mathbf{f}_{2}^{{}}\mathbf{f}_{2}^{\text{H}}\right)\text{vec}\left({{\mathbf{W}}^{\text{H}}}\right)={{\mathbf{w}}^{\text{H}}}\left(\mathbf{f}_{1}^{{}}\mathbf{f}_{1}^{\text{H}}\otimes\mathbf{f}_{2}^{{}}\mathbf{f}_{2}^{\text{H}}\right)\mathbf{w},
\end{aligned}
\end{equation}
where (a), (b), and (c) come from the fact that $\text{Tr}(\textbf{AB}) = \text{Tr}(\textbf{BA})$, $\text{Tr}(\textbf{A}^\text{H}\textbf{B}) = \text{vec}^{\text{H}}(\textbf{A}) \text{vec}(\textbf{B})$, and $\text{vec}(\textbf{ABC}) = (\textbf{C}^{\text{T}} \otimes \textbf{A})\text{vec}(\textbf{B})$, respectively. By the same token, we have,
\begin{equation}\label{f2Twf1power2}
{{\left|\mathbf{f}_{2}^{\text{T}}\mathbf{W}{{\mathbf{f}}_{1}}\right|}^{2}}={{\mathbf{w}}^{\text{H}}}\left(\mathbf{f}_{2}^{}\mathbf{f}_{2}^{\text{H}}\otimes\mathbf{f}_{1}^{{}}\mathbf{f}_{1}^{\text{H}}\right)\mathbf{w},\,
{{\|\mathbf{f}_{1}^{\text{T}}\mathbf{W}\|}^{2}}= {{\mathbf{w}}^{\text{H}}}\left(\mathbf{f}_{1}^{{}}\mathbf{f}_{1}^{\text{H}}\otimes\mathbf{I}\right)\mathbf{w},\,
{{\|\mathbf{f}_{2}^{\text{T}}\mathbf{W}\|}^{2}}= {{\mathbf{w}}^{\text{H}}}\left(\mathbf{f}_{2}^{{}}\mathbf{f}_{2}^{\text{H}}\otimes\mathbf{I}\right)\mathbf{w}.
\end{equation}
It is worth mentioning that as nodes $S_1$ and $S_2$ are of a single antenna, there is no proper way of preventing $E$ from eavesdropping during the first hop. However, in the second hop, the multi-antenna relay may send its transmitting signal vector in the null space of the eavesdropper's channel. The main objective of the current work is to find the best beamforming vector in the above mentioned null space such that the minimum achievable secrecy rate is maximized, improving fairness in such a network. As a result, the beamforming matrix at the relay should be adjusted such that the transmitted vector lies in the null space of the channel gain vector $\mathbf{f}_{e}$. Mathematically speaking, the following constraints should be applied,
\begin{equation}\label{null_1}
\mathbf{f}_{e}^{\text{T}}\mathbf{W}{{\mathbf{f}}_{1}}\overset{(\text{a})}{\mathop{=}}\,\text{Tr}\left(\mathbf{W}{{\mathbf{f}}_{1}}\mathbf{f}_{{e}}^{\text{T}}\right)
\overset{(b)}{\mathop{=}}\,\text{ve}{{\text{c}}^{\text{H}}}\left({{\mathbf{W}}^{\text{H}}}\right)\text{vec}\left({{\mathbf{f}}_{1}}\mathbf{f}_{{e}}^{\text{T}}\right)
={{\mathbf{w}}^{\text{H}}}\text{vec}\left({{\mathbf{f}}_{1}}\mathbf{f}_{{e}}^{\text{T}}\right)=0,
\end{equation}
\begin{equation}\label{null_2}
\mathbf{f}_{e}^{\text{T}}\mathbf{W}{{\mathbf{f}}_{2}}\overset{(\text{a})}{\mathop{=}}\,\text{Tr}\left(\mathbf{W}{{\mathbf{f}}_{2}}\mathbf{f}_{{e}}^{\text{T}}\right)
\overset{(b)}{\mathop{=}}\,\text{ve}{{\text{c}}^{\text{H}}}\left({{\mathbf{W}}^{\text{H}}}\right)\text{vec}\left({{\mathbf{f}}_{2}}\mathbf{f}_{{e}}^{\text{T}}\right)
={{\mathbf{w}}^{\text{H}}}\text{vec}\left({{\mathbf{f}}_{2}}\mathbf{f}_{{e}}^{\text{T}}\right)=0.
\end{equation}
where steps (\text{a}) and (\text{b}) of equations (\ref{null_1}) and (\ref{null_2}) are similar to steps (\text{a}) and (\text{b}) in (\ref{f1Twf2power2}).
On the other hand, constraints (\ref{null_1}) and (\ref{null_2}) can be combined into a single letter constraint as ${{\mathbf{w}}^{\text{H}}}\mathbf{Z}=\textbf{0}$ where $\mathbf{Z}\triangleq\left[ \begin{matrix}
\text{vec}\left( {{\mathbf{f}}_{1}}\mathbf{f}_{{e}}^{\text{T}} \right) & \text{vec}\left( {{\mathbf{f}}_{2}}\mathbf{f}_{{e}}^{\text{T}} \right)  \\ \end{matrix}\right]$. It should be noted that $\mathbf{Z}$ is at most rank 2. Hence, the rank ($m$) of its corresponding null space matrix $\mathbf{G}$ can take the value $m=N^2-2$ or $N^2-1$. Therefore, $\mathbf{w}$ can be chosen as $\mathbf{w}=\mathbf{G}\,\mathbf{c}$, where the matrix $\mathbf{G}_{N^2 \times m}$ has $m$ orthogonal eigen vectors corresponding to the zero singular values of the matrix ${{\mathbf{Z}}^{\text{H}}}$, and $\mathbf{c}_{m\times 1}$ is a combination vector, spanning the null space identified by $\mathbf{G}$. In other words, the columns of $\mathbf{G}$ subsume the basis vectors of the null space of $\textbf{Z}$ and $\mathbf{w}$ is written as a linear combination of these vectors. Using this approach, $N^2$ unknown beamforming variables are reduced to $m$ variables.

\begin{figure}[t]
\centering
    \subfigure[]{
         \label{sfig11}
         \includegraphics[scale=.25]{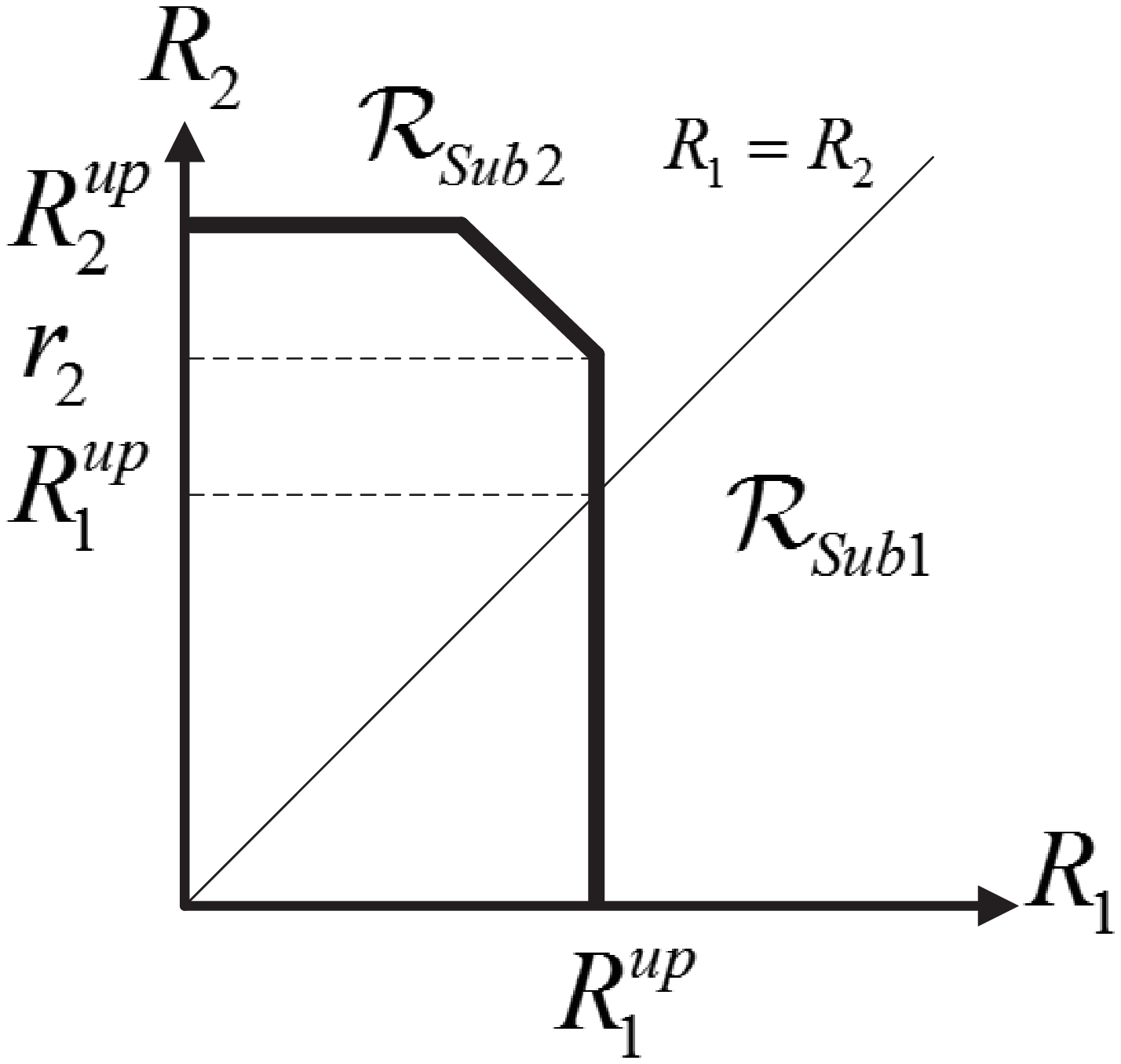}}
    \subfigure[]{
         \label{sfig12}
         \includegraphics[scale=.25]{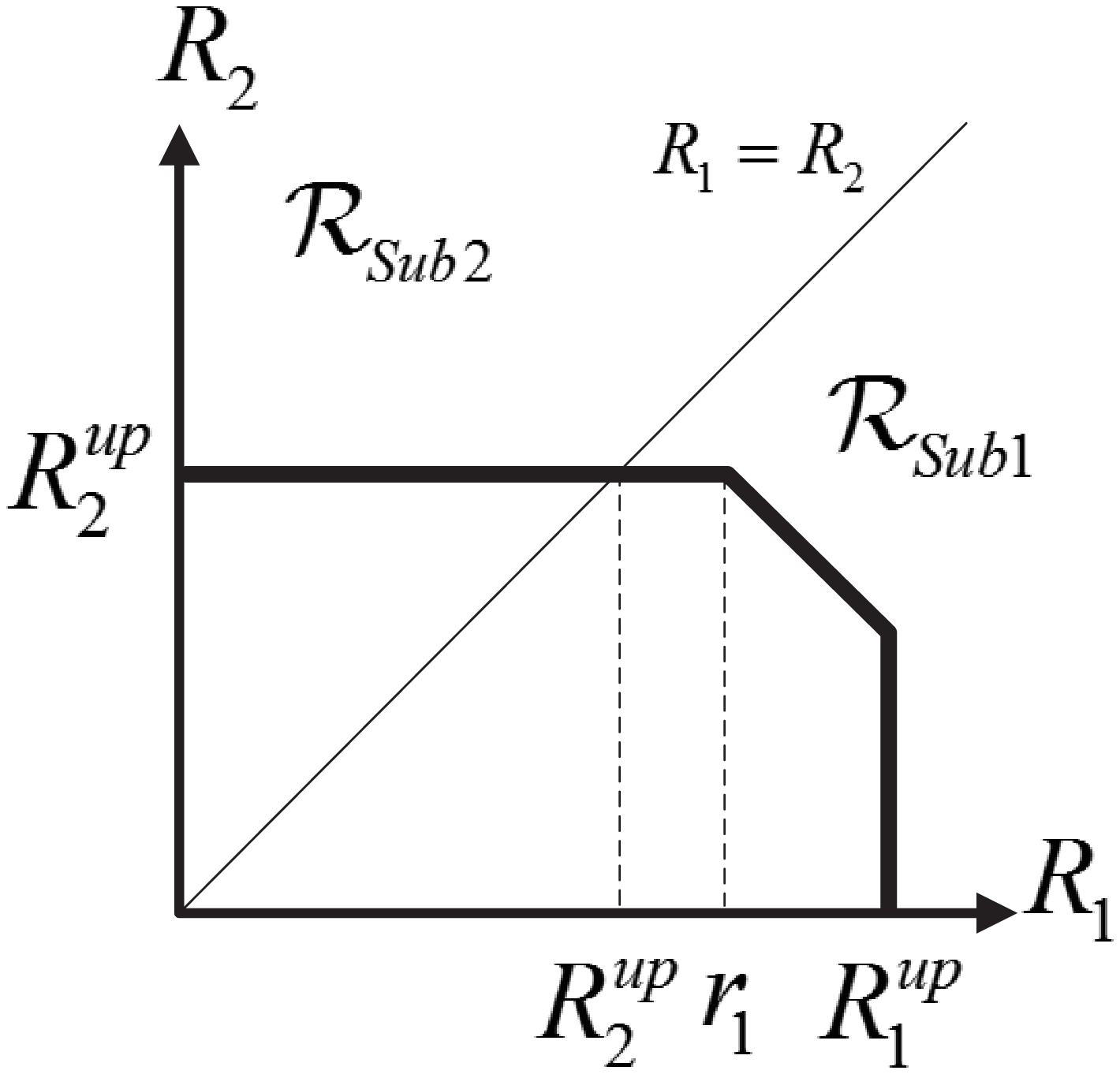}}\\
     \subfigure[]{
         \label{sfig13}
         \includegraphics[scale=.25]{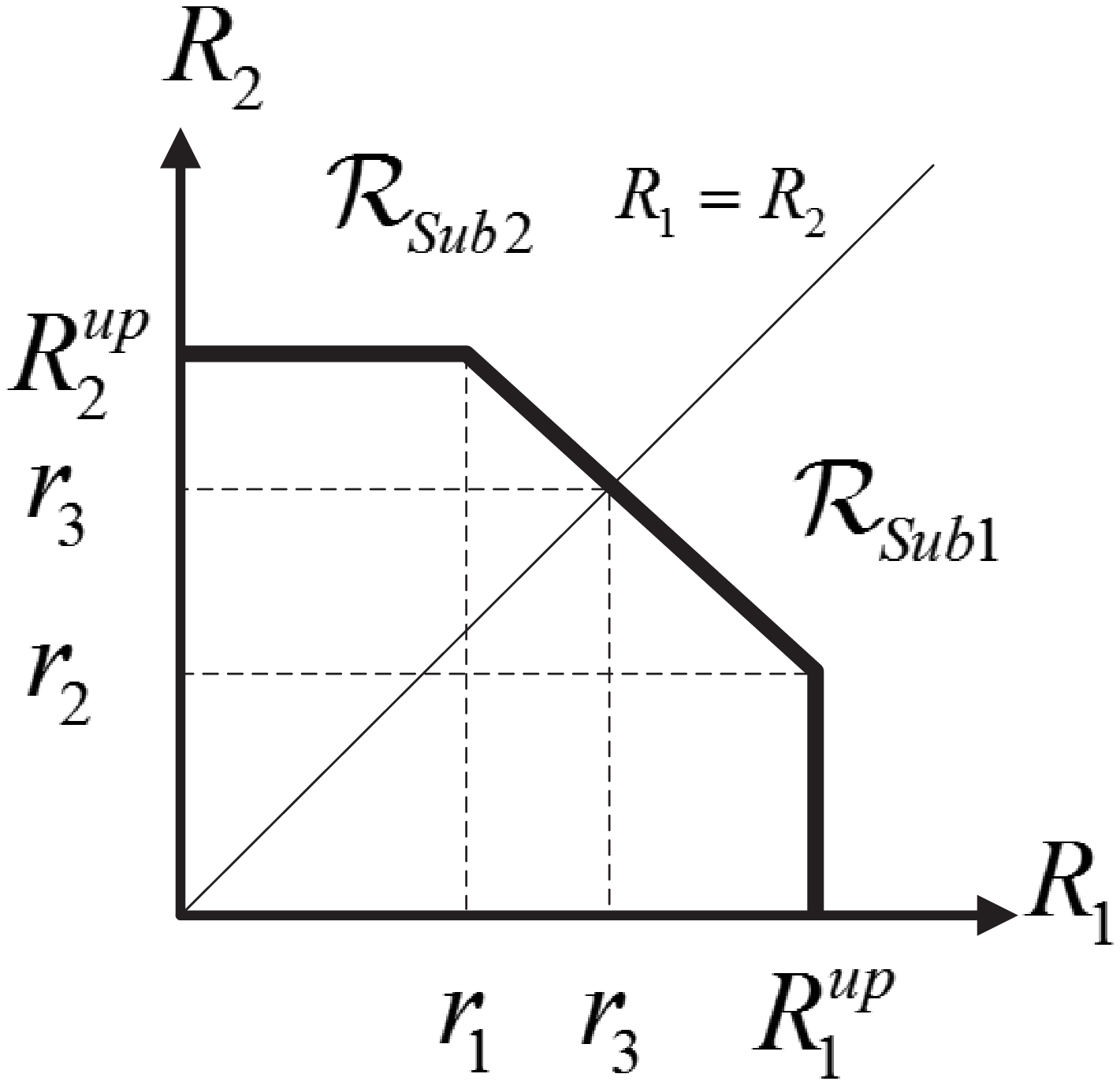}}
     \subfigure[]{
         \label{sfig14}
         \includegraphics[scale=.25]{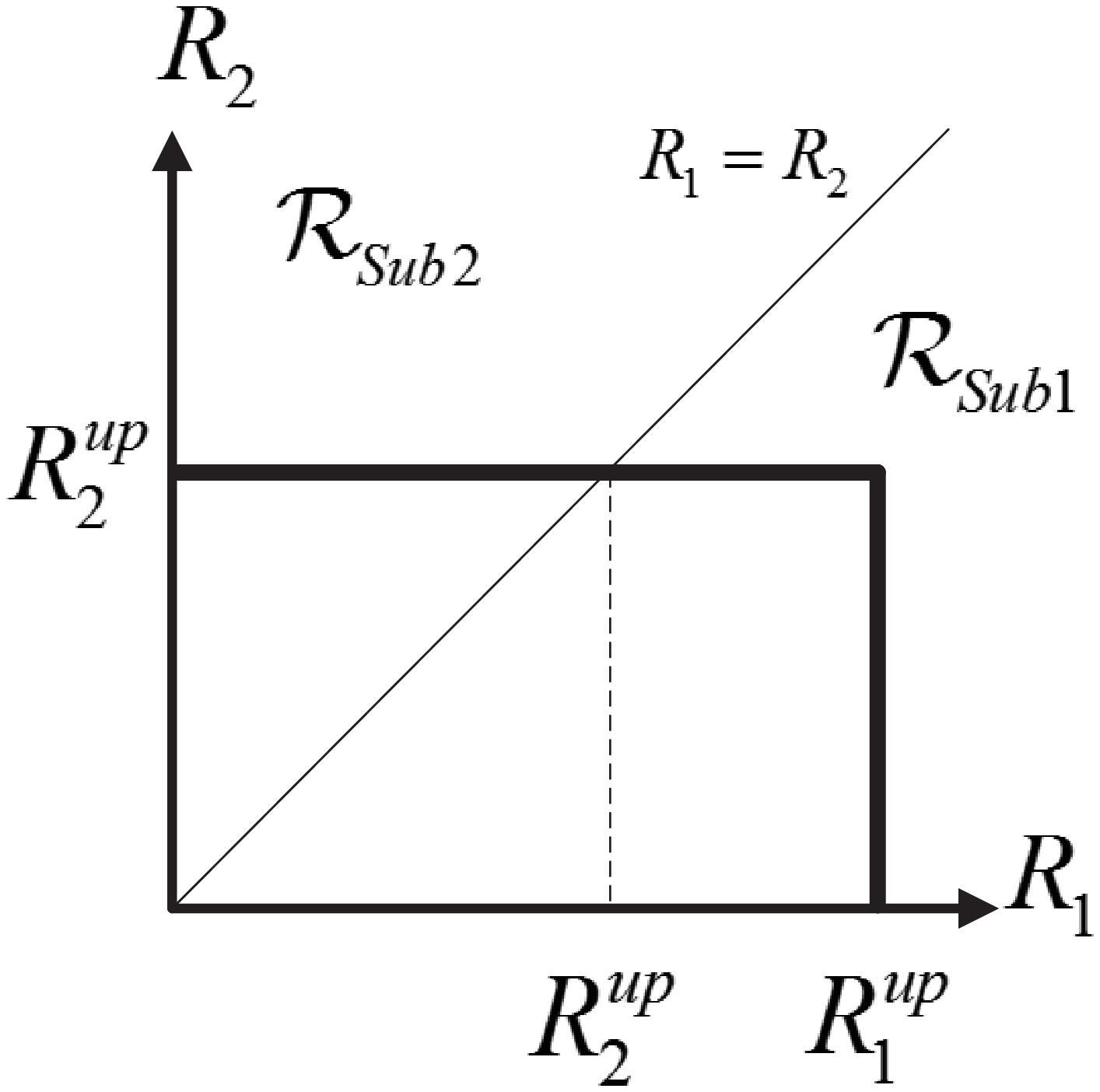}}
    \caption{The ASR region of a two-way channel : (a) when $R_{1}=R_{2}$ intersects $R_{1}=R_{1}^{up}$, (b) when $R_{1}=R_{2}$ intersects $R_{2}=R_{2}^{up}$, (c) when $R_{1}=R_{2}$ intersects $R_1+R_2=R_{sum}^{up}$,  (d) a special case of (a) or (b) }\label{fig1}
\end{figure}

Defining $R_{1}^{up} =\frac{1}{2}{{\left[ I\left( {{x}_{2}};{{y}_{1}} \right)-I\left( {{x}_{2}};{{\mathbf{y}}_{{E}}} \right) \right]}^{+}}$ and $R_{2}^{up} =\frac{1}{2}{{\left[ I\left( {{x}_{1}};{{y}_{2}} \right)-I\left( {{x}_{1}};{{\mathbf{y}}_{{E}}} \right) \right]}^{+}}$ as the upper bounds associated with the first two constraints in the ASR region introduced in (\ref{ASR Region}), using the equality $\mathbf{w} = \textbf{Gc}$ and the equations (\ref{Ix2y1}), (\ref{Ix1y2}), (\ref{botpage1}), (\ref{botpage2}), (\ref{f1Twf2power2})-(\ref{null_2}), and also after some mathematics it follows,
\begin{equation}\label{R_{1}^{*}}
R_{1}^{up}={{\left[ \frac{1}{2}{{\log }_{2}}\left( {{\tau }_{1}}+\frac{{{\mathbf{c}}^{\text{H}}}{{\mathbf{\Phi }}_{1}}\mathbf{c}}{{{\mathbf{c}}^{\text{H}}}{{\mathbf{\Sigma }}_{\text{1}}}\mathbf{c}+\sigma _{1}^{2}} \right) \right]}^{+}}
\end{equation}
and
\begin{equation}\label{R_{2}^{*}}
R_{2}^{up}={{\left[ \frac{1}{2}{{\log }_{2}}\left( {{\tau }_{2}}+\frac{{{\mathbf{c}}^{\text{H}}}{{\mathbf{\Phi }}_{2}}\mathbf{c}}{{{\mathbf{c}}^{\text{H}}}{{\mathbf{\Sigma }}_{2}}\mathbf{c}+\sigma _{2}^{2}} \right) \right]}^{+}},
\end{equation}
where the following definitions are being used,
\begin{equation}\label{defafterR1upandR2up}
\begin{aligned}
{{\tau }_{1}}\triangleq\left( 1+\frac{{{\text{P}}_{2}}{{\left| {{{g}}_{2}} \right|}^{2}}}{{{\text{P}}_{1}}{{\left| {{{g}}_{1}} \right|}^{2}}+\sigma _{{E,1}}^{2}} \right)^{-1},\,
{{\mathbf{\Phi }}_{1}}\triangleq{{\mathbf{G}}^{\text{H}}}{{\tau }_{1}}{{\text{P}}_{2}}\,\left( \mathbf{f}_{1}^{{}}\mathbf{f}_{1}^{\text{H}}\otimes \mathbf{f}_{2}^{{}}\mathbf{f}_{2}^{\text{H}} \right)\mathbf{G},\,
{{\mathbf{\Sigma }}_{\text{1}}}\triangleq{{\mathbf{G}}^{\text{H}}}\sigma _{{{R}}}^{2}\left( \mathbf{f}_{1}^{{}}\mathbf{f}_{1}^{\text{H}}\otimes \mathbf{I} \right)\mathbf{G},\\
{{\tau }_{2}}\triangleq\left( 1+\frac{{{\text{P}}_{1}}{{\left| {{{g}}_{1}} \right|}^{2}}}{{{\text{P}}_{2}}{{\left| {{{g}}_{2}} \right|}^{2}}+\sigma _{{E,1}}^{2}} \right)^{-1},\,
{{\mathbf{\Phi }}_{2}}\triangleq{{\mathbf{G}}^{\text{H}}}{{\tau }_{2}}{{\text{P}}_{1}}\,\left( \mathbf{f}_{2}^{{}}\mathbf{f}_{2}^{\text{H}}\otimes \mathbf{f}_{1}^{{}}\mathbf{f}_{1}^{\text{H}} \right)\mathbf{G},\,
{{\mathbf{\Sigma }}_{2}}\triangleq{{\mathbf{G}}^{\text{H}}}\sigma _{{{R}}}^{2}\left( \mathbf{f}_{2}^{{}}\mathbf{f}_{2}^{\text{H}}\otimes \mathbf{I} \right)\mathbf{G}.
\end{aligned}
\end{equation}
Also, noting (\ref{ASR Region}), the upper bound for the sum secrecy rate, defined as $R_{sum}^{up}$, can be computed as,
 \begin{equation}\label{Rsumupmutualinf}
R_{sum}^{up}=\frac{1}{2}{{\left[ I\left( {{{x}}_{1}};{{{y}}_{2}} \right)+I\left( {{x}_{2}};{{y}_{1}} \right)-I\left( {{{x}}_{1}}{,}{{{x}}_{2}};{{\mathbf{y}}_{{E}}} \right) \right]}^{+}},
\end{equation}
where, using (\ref{Ix2y1})-(\ref{Ix1x2YE}), and (\ref{f1Twf2power2})-(\ref{null_2}), $R_{sum}^{up}$ can be simplified to,
\begin{equation}\label{e30}
\begin{aligned}
{{R}_{sum}^{up}}=
{{\left[ \frac{1}{2}{{\log }_{2}}\delta \left( 1+\frac{{{\mathbf{c}}^{\text{H}}}{{\mathbf{\Phi }}_{31}}\mathbf{c}}{{{\mathbf{c}}^{\text{H}}}{{\mathbf{\Sigma }}_{\text{1}}}\mathbf{c}+\sigma _{1}^{2}}\right) \left( 1+\frac{{{\mathbf{c}}^{\text{H}}}{{\mathbf{\Phi }}_{32}}\mathbf{c}}{{{\mathbf{c}}^{\text{H}}}{{\mathbf{\Sigma }}_{\text{2}}}\mathbf{c}+\sigma _{2}^{2}}\right) \right]}^{+}},
\end{aligned}
\end{equation}
where the following definitions are being used,
\begin{equation}\label{DdeltaA}
\begin{aligned}
{\delta }\triangleq \left(1+\frac{{{\text{P}}_{1}}{{\left| {{{g}}_{1}}\right|}^{2}}+{{\text{P}}_{2}}{{\left| {{{g}}_{2}} \right|}^{2}}}{\sigma _{{E,1}}^{2}}\right)^{-1},\,
{{\mathbf{\Phi }}_{31}}\triangleq \frac{{{\mathbf{\Phi }}_{1}}}{\tau_1} ,\,
{{\mathbf{\Phi }}_{32}}\triangleq \frac{{{\mathbf{\Phi }}_{2}}}{\tau_2}.
\end{aligned}
\end{equation}
Fig. \ref{fig1} shows the ASR region for different cases of constraints w.r.t. the line $R_1=R_2$. Based on different channel conditions and the chosen beamforming matrix at the relay, the line $R_{1}=R_{2}$ intersects one of the border lines $R_{1}=R_{1}^{up}$, $R_{2}=R_{2}^{up}$, or $R_1+R_2=R_{sum}^{up}$ which are depicted in Figs. \ref{sfig11}-\ref{sfig13}, respectively. Fig. \ref{sfig14} indicates the case in which the third constraint of (\ref{ASR Region}) is overlaid with the other two constraints, thus it is a redundant constraint, not to be taken into account. This case only happens when at least one of the channel gains $g_1$ or $g_2$ takes the zero value. This case represents a rectangular ASR region and can be considered as a special case of Fig. \ref{sfig11} or Fig. \ref{sfig12}.

According to Fig. \ref{sfig11}, the point $(R_1^{up},{r}_{2})$ identifies the intersection point of lines $R_1 = R_1^{up}$ and $R_1+R_2 = R_{sum}^{up}$, therefore referring to (\ref{Ix2y1})-(\ref{Ix1x2YE}), (\ref{f1Twf2power2})-(\ref{null_2}) and (\ref{R_{1}^{*}}), and using the equation $R_{1}^{up}\text{+}{{r}_{2}}=
\frac{1}{2}{{\left[ {I}\left( {{{x}}_{1}};{{{y}}_{2}} \right)+{I}\left( {{{x}}_{2}};{{{y}}_{1}} \right)-{I}\left( {{{x}}_{1}}\text{,}{{{x}}_{2}};{{\mathbf{y}}_{{E}}} \right) \right]}^{+}}$, $r_2$ can be computed as,
\begin{equation}\label{e33}
{{r}_{2}}=\frac{1}{2}{{\log }_{2}}\left( {{\tau }_{3}}+\frac{{{\mathbf{c}}^{\text{H}}}{{\mathbf{\Omega }}_{1}}\mathbf{c}}{{{\mathbf{c}}^{\text{H}}}{{\mathbf{\Sigma }}_{\text{2}}}\mathbf{c}+\sigma _{2}^{2}} \right),
\end{equation}
where ${{\tau }_{3}}\triangleq \frac{\delta}{\tau_1} $ and ${{\mathbf{\Omega }}_{1}}\triangleq{{\tau }_{3}}{{\mathbf{\Phi }}_{32}}$. Similarly, $({r}_{1},R_2^{up})$ is the intersection point of the lines $R_2 = R_2^{up}$ and $R_1+R_2 = R_{sum}^{up}$. Therefore, using (\ref{Ix2y1})-(\ref{Ix1x2YE}), (\ref{f1Twf2power2})-(\ref{null_2}), and (\ref{R_{2}^{*}}), one can write
\begin{equation}\label{r1}
{{r}_{1}}=\frac{1}{2}{{\log }_{2}}\left( {{\tau }_{4}}+\frac{{{\mathbf{c}}^{\text{H}}}{{\mathbf{\Omega }}_{2}}\mathbf{c}}{{{\mathbf{c}}^{\text{H}}}{{\mathbf{\Sigma }}_{\text{1}}}\mathbf{c}+\sigma _{1}^{2}} \right),
\end{equation}
where ${{\tau }_{4}}\triangleq \frac{\delta}{\tau_2} $ and ${{\mathbf{\Omega }}_{2}}\triangleq {{\tau }_{4}} {{\mathbf{\Phi }}_{31}} $. Also, $({r}_{3},{r}_{3})$ is the intersection point of the lines $R_1 = R_2$ and $R_1+R_2 = R_{sum}^{up}$. Thus by considering (\ref{e30}), $r_3$ can be calculated as,
\begin{equation}\label{e39}
\begin{aligned}
{{r}_{3}}=\frac{1}{2}{R}_{sum}^{up} =
{\left[ \frac{1}{4}{{\log }_{2}}\delta \left( 1+\frac{{{\mathbf{c}}^{\text{H}}}{{\mathbf{\Phi }}_{31}}\mathbf{c}}{{{\mathbf{c}}^{\text{H}}}{{\mathbf{\Sigma }}_{\text{1}}}\mathbf{c}+\sigma _{1}^{2}}\right) \left( 1+\frac{{{\mathbf{c}}^{\text{H}}}{{\mathbf{\Phi }}_{32}}\mathbf{c}}{{{\mathbf{c}}^{\text{H}}}{{\mathbf{\Sigma }}_{\text{2}}}\mathbf{c}+\sigma _{2}^{2}}\right) \right]}^{+}.
\end{aligned}
\end{equation}

\section{PROBLEM FORMULATION}\label{sec:prob}
This section aims to find the best NSBF matrix at the relay, i.e., $\textbf{W}$ or equivalently $\mathbf{w} \triangleq \text{vec}(\textbf{W}^\text{H})=\mathbf{G}\bold{c}$, such that the minimum ASR of two nodes is maximized, assuming the relay is subject to a peak power constraint. Mathematically speaking, since knowing $\mathbf{c}$ leads to the $\mathbf{w}$, the following optimization problem should be addressed,
\begin{equation}\label{main_problem}
    \begin{aligned}
    \underset{\mathcal{R},\mathbf{c}}{\mathop{\text{max}}}\,\,\text{min}\left( {{R}_{1}},{{R}_{2}} \right)
    \,\text{s.t.}\,\,\,\{{{\text{p}}_{{R}}}\le {{\text{P}}_{{R}}}\},
    \end{aligned}
\end{equation}
where $\mathcal{R}$ denotes the ASR region determined by (\ref{ASR Region}) for a known value of $\mathbf{c}$, ${{R}_{1}}$ and ${{R}_{2}}$ are the ASRs of two transmitting ends, ${{\text{P}}_{{R}}}$ is the total available power at the relay and ${{\text{p}}_{{R}}}$ is the
relay transmit power, where noting the relay's transmit signal vector is $\mathbf{s}=\mathbf{Wr}$, the transmit power can be calculated as,
\begin{equation}\label{e41}
\begin{aligned}
{{\text{p}}_{{R}}}&=\text{E}\left\{ {{\mathbf{s}}^{\text{H}}}\mathbf{s} \right\}= \text{Tr}\left( \mathbf{W}\text{E}\left\{{{\mathbf{rr^\text{H}}}}\right\}{{\mathbf{W}}^{\text{H}}} \right)
=\,\text{Tr}\left( \mathbf{W}{{\mathbf{Q}}_{\mathbf{r}}}{{\mathbf{W}}^{\text{H}}} \right)\overset{\left( \text{a} \right)}{\mathop{=}}\,\text{vec}{{\left( \mathbf{Q}_{\mathbf{r}}^{\text{T}}\mathbf{I}{{\mathbf{W}}^{\text{T}}} \right)}^{\text{T}}}\text{vec}\left( {{\mathbf{W}}^{\text{H}}} \right)\\
&\overset{\left( \text{b} \right)}{\mathop{=}}\,\text{vec}{{\left( \mathbf{I} \right)}^{\text{T}}}{{\left( \mathbf{W}\otimes \mathbf{Q}_{\mathbf{r}}^{\text{T}} \right)}^{\text{T}}}\text{vec}\left( {{\mathbf{W}}^{\text{H}}} \right)
\overset{\left( \text{c} \right)}{\mathop{=}}\,\text{vec}{{\left( \mathbf{I} \right)}^{\text{T}}}\left( {{\mathbf{W}}^{\text{T}}}\otimes \mathbf{Q}_{\mathbf{r}}^{{}} \right)\text{vec}\left( {{\mathbf{W}}^{\text{H}}} \right)\\
&\overset{\left( \text{d} \right)}{\mathop{=}}\,\text{vec}{{\left( {{\mathbf{W}}^{\text{T}}} \right)}^{\text{T}}}\left( \mathbf{I}\otimes \mathbf{Q}_{\mathbf{r}}^{{}} \right)\text{vec}\left( {{\mathbf{W}}^{\text{H}}} \right)
\overset{\left( \text{e} \right)}{\mathop{=}}\,{{\mathbf{w}}^{\text{H}}}\left( \mathbf{I}\otimes \mathbf{Q}_{\mathbf{r}}^{{}} \right)\mathbf{w}
={{\mathbf{c}}^{\text{H}}}{{\mathbf{\Omega }}_{{R}}}\mathbf{c},
\end{aligned}
\end{equation}
where in (\ref{e41}) (a), (b), (c), and (d) come from $\text{Tr}\left( {{\mathbf{A}}^{\text{T}}}\mathbf{B} \right)=\text{vec}{{\left( \mathbf{A} \right)}^{\text{T}}}\text{vec}\left( \mathbf{B} \right)$, $\text{vec}(\textbf{ABC}) = (\textbf{C}^{\text{T}} \otimes \textbf{A})\text{vec}(\textbf{B})$, ${{\left( \mathbf{A}\otimes \mathbf{B} \right)}^{\text{T}}}=\left( {{\mathbf{A}}^{\text{T}}}\otimes {{\mathbf{B}}^{\text{T}}} \right)$, and $\text{vec}{{\left( \mathbf{I} \right)}^{\text{T}}}\left( {{\mathbf{A}}^{\text{T}}}\otimes \mathbf{B} \right)=\text{vec}{{\left( {{\mathbf{A}}^{\text{T}}} \right)}^{\text{T}}}\left( \mathbf{I}\otimes \mathbf{B} \right)$, respectively. Moreover, the definition $\mathbf{w}=\text{vec}\left( {{\mathbf{W}}^{\text{H}}} \right)$ gives ${{\mathbf{w}}^{\text{H}}}=\text{vec}{{\left( {{\mathbf{W}}^{\text{T}}} \right)}^{\text{T}}}$ leading to the equality (e). Also, $\overset{{}}{\mathop{{{\mathbf{\Omega }}_{{R}}}\triangleq}}\,{{\mathbf{G}}^{\text{H}}}\left( \mathbf{I}\otimes \mathbf{Q}_{\text{r}}^{{}} \right)\mathbf{G}$ and ${{\mathbf{Q}}_{\mathbf{r}}}=\text{E}\left\{ {{\mathbf{r}}}\mathbf{r}^{\text{H}} \right\}
={{\text{P}}_{1}}\mathbf{f}_{1}^{{}}\mathbf{f}_{1}^{\text{H}}+{{\text{P}}_{2}}\mathbf{f}_{2}^{{}}\mathbf{f}_{2}^{\text{H}}+\sigma_{{R}}^{2}{{\mathbf{I}}}$.

The optimization problem in (\ref{main_problem}) can be rewritten as
\begin{equation}\label{e43}
    \begin{aligned}
    \underset{\mathbf{c}}{\mathop{\text{max}}}\,Q\left( \mathbf{c} \right)
    \,\text{s}\text{.t.}\,\,\,\{{{\text{p}}_{{R}}}\le {{\text{P}}_{{R}}}\},
    \end{aligned}
\end{equation}
where $Q\left( \mathbf{c} \right)=\underset{\mathcal{R}}{\mathop{\text{max}}}\,\min\left( {{R}_{1}},{{R}_{2}} \right)$.

As is noted in the preceding section, Fig. \ref{fig1} illustrates different possible choices of region $\mathcal{R}$ w.r.t. the line $R_1=R_2$. In each shape, the region $\mathcal{R}$ is divided into two sub-regions $\mathcal{R}_{Sub1}$ and $\mathcal{R}_{Sub2}$ by the line $R_1=R_2$. In the sub-region $\mathcal{R}_{Sub1}$, the value of $R_{2}$ is less than or equal $R_{1}$ in all points, thus, we have $\min(R_1,R_2)=R_2$, where the maximum value of $R_{2}$ occurs on the intersection of the line $R_1=R_2$ with the boundary of $\mathcal{R}_{Sub1}$. Similarly, in the sub-region $\mathcal{R}_{Sub2}$, we have $\min(R_1,R_2)=R_1$, where again the maximum value of $R_1$ resides on the boundary of this sub-region.

In Fig. \ref{fig1}, the sub-figure (d) can be considered as a special case of sub-figures (a) or (b). So, it is adequate to merely study sub-figures (a)-(c). It should be noted that for each value of $\textbf{c}$, one of these regions occurs. Noting this, one way to tackle the problem is to divide the feasible set of $\textbf{c}$ into three sub-sets, each corresponding to one of sub-figures (a)-(c). Using this approach, the main problem (\ref{main_problem}) is replaced by three sub-problems with three different sub-sets, where the one that yields the maximum value of $\min(R_1,R_2)$ would be the optimal solution of (\ref{main_problem}). For instance, in sub-figure (a) we have $\text{max} ~\text{min} \{ R_1,R_2 \} = R_1^{up}$, where according to Fig. \ref{fig1}~(a) this sub-figure occurs if and only if we have $R_1^{up}\leq r_2$ . Thus, the corresponding sub-problem can be written as,
\begin{equation}\label{sub_1}
    \begin{aligned}
    \underset{\mathbf{c}_1}{\mathop{\text{max}}}\,R_{1}^{up}
    \,\,\,\,\text{s}\text{.t.}\,\,\,\{{{\text{p}}_{{R}}}\le {{\text{P}}_{{R}}}
    ,\,R_{1}^{up}\le {{r}_{2}}\},
    \end{aligned}
\end{equation}
where the second constraint ensures that sub-figure (a) occurs.
Similarly, the second and the third sub-problems are given, respectively, by
\begin{equation}\label{sub_2}
    \begin{aligned}
    \underset{\mathbf{c}_2}{\mathop{\text{max}}}\,R_{2}^{up}
    \,\,\,\,\text{s}\text{.t}\text{.}\,\,\,\,\{{{\text{p}}_{{R}}}\le {{\text{P}}_{{R}}}
    ,\,R_{2}^{up}\le {{r}_{1}}\}
    \end{aligned}
\end{equation}
and
\begin{equation}\label{sub_3}
    \begin{aligned}
    \underset{\mathbf{c}_3}{\mathop{\text{max}}}\,\,\,\,{r}_{3}
    \,\,\,\,\text{s}\text{.t.}\,\,\,\,\{{{\text{p}}_{{R}}}\le {{\text{P}}_{{R}}}
    ,\,R_{1}^{up}>{{r}_{2}}
    ,\,R_{2}^{up}>{{r}_{1}}\}.
    \end{aligned}
\end{equation}
Using the values of $R_{1}^{up}$ in (\ref{R_{1}^{*}}), ${r}_2$ in (\ref{e33}), and ${{\text{p}}_{{R}}}$ in (\ref{e41}), also noting the logarithm function is monotonically increasing and ${{\tau }_{1}}$ is a positive constant value, the sub-problem (\ref{sub_1}) can be rewritten as,
\begin{equation}\label{p1ellog}
\begin{aligned}
&\underset{\mathbf{c}_1}{\mathop{\text{max}}}\,\,\frac{{{\mathbf{c}_1}^{\text{H}}}{{\mathbf{\Phi }}_{1}}\mathbf{c}_1}{{{\mathbf{c}_1}^{\text{H}}}{{\mathbf{\Sigma }}_{\text{1}}}\mathbf{c}_1+\sigma _{1}^{2}}\,\,\,\,
\text{s}\text{.t.}~~~\{{{\mathbf{c}_1}^{\text{H}}}{{\mathbf{\Omega }}_{{R}}}\mathbf{c}_1\le {{\text{P}}_{{R}}},
\,\,\,\,\frac{{{\mathbf{c}_1}^{\text{H}}}{{\mathbf{\Phi }}_{1}}\mathbf{c}_1}{{{\mathbf{c}_1}^{\text{H}}}{{\mathbf{\Sigma }}_{\text{1}}}\mathbf{c}_1+\sigma _{1}^{2}}-\frac{{{\mathbf{c}_1}^{\text{H}}}{{\mathbf{\Omega }}_{1}}\mathbf{c}_1}{{{\mathbf{c}_1}^{\text{H}}}{{\mathbf{\Sigma }}_{\text{2}}}\mathbf{c}_1+\sigma _{2}^{2}}\le {{\tau }_{3}}-{{\tau }_{1}}\}.
\end{aligned}
\end{equation}
The problem (\ref{p1ellog}) is not convex in general, thus it does not have a trivial solution. To simplify this problem, the SDR method is employed in the current work. To this end, one can define $\mathbf{C}_1=\mathbf{c}_1\mathbf{c}_1^\text{H}$  and rewrite the problem (\ref{p1ellog}) as:
\begin{equation}\label{p1Trform}
    \begin{aligned}
        \underset{\mathbf{C}_1}{\mathop{\text{max}}}\,\,\frac{\text{Tr}\left( {{\mathbf{\Phi }}_{1}}\mathbf{C}_1 \right)}{\text{Tr}\left( {{\mathbf{\Sigma }}_{\text{1}}}\mathbf{C}_1 \right)+\sigma _{1}^{2}}
         \,\,\,\,\text{s}\text{.t.}&\,\,\,\,\{\text{Tr}\left( {{\mathbf{\Omega }}_{{R}}}\mathbf{C}_1 \right)\le {{\text{P}}_{{R}}}, \\
         &\frac{\text{Tr}\left({{\mathbf{\Phi}}_{1}}\mathbf{C}_1 \right)}{\text{Tr}\left( {{\mathbf{\Sigma }}_{\text{1}}}\mathbf{C}_1 \right)+\sigma _{1}^{2}}-\frac{\text{Tr}\left( {{\mathbf{\Omega }}_{1}}\mathbf{C}_1 \right)}{\text{Tr}\left( {{\mathbf{\Sigma }}_{\text{2}}}\mathbf{C}_1 \right)+\sigma _{2}^{2}}\le {{\tau }_{3}}-{{\tau}_{1}},
         \,\,\,\mathbf{C}_1\succeq0,\,\, \text{rank}\left( \mathbf{C}_1 \right)=1\},
    \end{aligned}
\end{equation}
where the rank-one constraint is non-convex, and according to the SDR method, it should be discarded. Furthermore, the objective function of (\ref{p1Trform}) can be simplified using the Charnes-Cooper transformation \cite{Charnescooper}. To this end, the scalar $\zeta_1>0$ and the matrix $\mathbf{Z}_1$ are defined such that $\textbf{C}_1 = \frac{\textbf{Z}_1}{\zeta_1}$. Using this approach, $\textbf{Z}_1$ and $\zeta_1$ should be computed such that the denominator of the objective function of (\ref{p1Trform}) times $\zeta_1$ takes unit value. As a result, the problem (\ref{p1Trform}) is changed to,
\begin{equation}\label{subp1aftercooper}
    \begin{aligned}
        \underset{{{\mathbf{Z}}_{1}}\,,\,{{\zeta }_{1}}\,}{\mathop{\max }}\,\,\,\,\text{Tr}\left( {{\mathbf{\Phi }}_{1}}{{\mathbf{Z}}_{1}} \right)
        \,\,\,\, \text{s}\text{.t}\text{.}\,\,\,\,&\{\text{Tr}\left( {{\mathbf{\Omega }}_{{R}}}{{\mathbf{Z}}_{1}} \right)\le {{\zeta }_{1}}{{\text{P}}_{R}},\,\,\,\,
        \text{Tr}\left( {{\mathbf{\Sigma }}_{\text{1}}}{{\mathbf{Z}}_{1}} \right)+{{\zeta}_{1}}\sigma _{1}^{2}=1,\\
        &\text{Tr}\left( {{\mathbf{\Phi}}_{1}}{{\mathbf{Z}}_{1}}\right)-\frac{\text{Tr}\left( {{\mathbf{\Omega}}_{1}}{{\mathbf{Z}}_{1}}\right)}{\text{Tr}\left( {{\mathbf{\Sigma }}_{\text{2}}}{{\mathbf{Z}}_{1}} \right)+{{\zeta }_{1}}\sigma _{2}^{2}}\le {{\tau}_{3}}-{{\tau }_{1}},\,\,\,\,{{\zeta }_{1}}>0,\,\,\,\,{\mathbf{Z}}_{1}\succeq0\}.
    \end{aligned}
\end{equation}
The third constraint in (\ref{subp1aftercooper}) is not convex in general. To address this issue, it is assumed that the term ${\text{Tr}\left( {{\mathbf{\Sigma }}_{\text{2}}}{{\mathbf{Z}}_{1}} \right)+{{\zeta }_{1}}\sigma _{2}^{2}}$ is a constant value and equals $t_1$. Therefore, this problem can be simplified to,
\begin{equation}\label{subp1aftercooper2}
    \begin{aligned}
    \underset{{{\mathbf{Z}}_{1}}\,,\,{{\zeta }_{1}}\,}{\mathop{\max }}\,\,\,\,\text{Tr}\left( {{\mathbf{\Phi}}_{1}}{{\mathbf{Z}}_{1}}\right)\,\,\,\,
     \text{s}\text{.t}\text{.}\,\,\,\,&\{\text{Tr}\left( {{\mathbf{\Omega }}_{{R}}}{{\mathbf{Z}}_{1}} \right)\le {{\zeta }_{1}}{{\text{P}}_{R}},
     \,\,\,\,\text{Tr}\left( {{\mathbf{\Sigma }}_{\text{1}}}{{\mathbf{Z}}_{1}} \right)+{{\zeta }_{1}}\sigma _{2}^{2}=1, \\
     &\text{Tr}\left( {{\mathbf{\Phi }}_{1}}{{\mathbf{Z}}_{1}} \right)-\frac{\text{Tr}\left( {{\mathbf{\Omega }}_{1}}{{\mathbf{Z}}_{1}} \right)}{{{t}}_{1}}\le {{\tau }_{3}}-{{\tau }_{1}},\,\,\,\,
      \text{Tr}\left( {{\mathbf{\Sigma }}_{\text{2}}}{{\mathbf{Z}}_{1}} \right)+{{\zeta }_{1}}\sigma _{2}^{2}\text{ = }{{t}_{1}},\\
       &{{\zeta }_{1}}>0,~~{\mathbf{Z}_1}\succeq0\}.
    \end{aligned}
\end{equation}
As seen, knowing $t_1$, the third constraint of (\ref{subp1aftercooper2}) is affine, hence the problem (\ref{subp1aftercooper2}) is an SDP problem. This problem can be numerically solved using the interior point methods embedded in SDP problem solvers, including the CVX package, which is employed in the current study. Note that a one-dimensional search should be performed on $t_1$, where the problem (\ref{subp1aftercooper2}) should be tackled for each value of $t_1$. Finally, the best answer is chosen as the solution of (\ref{subp1aftercooper}). In the Appendix, an upper bound is provided for $t_1$ to restrict the search process in a bounded interval. The following remark discusses the optimal/suboptimal solution of the original problem in (\ref{p1ellog}).
\begin{remark}\label{remark1}
  In (\ref{subp1aftercooper2}), the best values of $\textbf{Z}_1$ and $\zeta_1$ are denoted by $\textbf{Z}_1^*$ and $\zeta_1^*$, respectively. If $\textbf{Z}_1^*$ is of rank one, the optimum value of $\textbf{C}_{1}$ in (\ref{p1Trform}) is given by $\textbf{C}_1^* = \frac{\textbf{Z}_1^*}{\zeta_1^*}\,$. In this case, the optimal value of $\textbf{c}$ in (\ref{p1ellog}) is the principle eigenvector of $\textbf{C}_1^*$. Otherwise, the principal eigenvector of the matrix $\textbf{C}_1 = \frac{\textbf{Z}_1^*}{\zeta_1^*}$ is a suboptimal solution, and the objective function of (\ref{p1Trform}) can be considered as an upper bound of (\ref{p1ellog}).
\end{remark}

Substituting $R_{2}^{up}$, ${r}_1$, ${{\text{p}}_{{R}}}$, respectively, from (\ref{R_{2}^{*}}), (\ref{r1}), and (\ref{e41}) in the sub-problem (\ref{sub_2}), a mathematically similar problem to (\ref{p1ellog}) is obtained; thus, employing the same approach, the following optimization problem should be tackled for a known value of $t_2$.
\begin{equation}\label{subp2aftercooper}
    \begin{aligned}
          \underset{{{\mathbf{Z}}_{2}}\,,\,{{\zeta }_{2}}\,}{\mathop{\max }}\,\,\,\,\text{Tr}\left( {{\mathbf{\Phi }}_{2}}{{\mathbf{Z}}_{2}} \right) \,\,\,\,\text{s}\text{.t.}\,\,\,\,&\{\text{Tr}\left( {{\mathbf{\Omega }}_{{R}}}{{\mathbf{Z}}_{2}} \right)\le {{\zeta }_{2}}{{\text{P}}_{R}},\,\,\,\,
         \text{Tr}\left( {{\mathbf{\Sigma }}_{2}}{{\mathbf{Z}}_{2}} \right)+{{\zeta }_{2}}\sigma _{2}^{2}=1,\\ &\text{Tr}\left( {{\mathbf{\Phi }}_{2}}{{\mathbf{Z}}_{2}} \right)-\frac{\text{Tr}\left( {{\mathbf{\Omega }}_{2}}{{\mathbf{Z}}_{2}} \right)}{{{t}_{2}}}\le {{\tau }_{4}}-{{\tau }_{2}},\,\,\,\,\text{Tr}\left( {{\mathbf{\Sigma }}_{1}}{{\mathbf{Z}}_{2}} \right)+{{\zeta }_{2}}\sigma _{1}^{2}\text{ = }{{t}_{2}}, \\
         & \,{{\zeta }_{2}}>0,~~\mathbf{\mathbf{Z}}_2\succeq0\}.
    \end{aligned}
\end{equation}
Since a one-dimensional search for the best value of $t_2$ should be done to get the optimal value, using the feasibility study, an upper bound for $t_2$ is determined in the Appendix, limiting the search space.

Also, substituting ${r}_1$, $r_2$, $r_3$, $R_1^{up}$, $R_2^{up}$, and ${{\text{p}}_{{R}}}$, respectively, from (\ref{r1}), (\ref{e33}), (\ref{e39}), (\ref{R_{1}^{*}}), (\ref{R_{2}^{*}}), and (\ref{e41}) in the sub-problem (\ref{sub_3}), and defining the slack variable $t_3=\frac{{{\mathbf{c}_3}^{\text{H}}} {{\mathbf{\Phi }}_{32}} {{\mathbf{c}_3}}}{{{\mathbf{c}_3}^{\text{H}}} {{\mathbf{\Sigma }}_{\text{2}}} {{\mathbf{c}_3}} + \sigma _{2}^{2}}$, it follows,
\begin{equation}\label{subp32}
    \begin{aligned}
    \underset{\mathbf{c}_3,t_3}{\mathop{\text{max}}}\,&\left( 1+\frac{{{\mathbf{c}_3}^{\text{H}}}{{\mathbf{\Phi }}_{31}}\mathbf{c}_3}{{{\mathbf{c}_3}^{\text{H}}}{{\mathbf{\Sigma }}_{\text{1}}}\mathbf{c}_3+\sigma _{1}^{2}} \right)\left( 1+{{t}_{3}} \right)
    \,\,\,\,\text{s}\text{.t.}\,\,\,\,\{{{\mathbf{c}_3}^{\text{H}}}{{\mathbf{\Omega }}_{{R}}}\mathbf{c}_3\le {{\text{P}}_{{R}}},\,\,\,\,
    {{\mathbf{c}_3}^{\text{H}}}\left( {{\mathbf{\Phi }}_{32}}-{{t}_{3}}{{\mathbf{\Sigma }}_{\text{2}}} \right)\mathbf{c}_3 = {{t}_{3}}\sigma _{2}^{2}, \\
      &\frac{{{\mathbf{c}_3}^{\text{H}}}{{\mathbf{\Phi }}_{1}}\mathbf{c}_3}{{{\mathbf{c}_3}^{\text{H}}}{{\mathbf{\Sigma }}_{\text{1}}}\mathbf{c}_3+\sigma _{1}^{2}}-\frac{{{\mathbf{c}_3}^{\text{H}}}{{\mathbf{\Omega }}_{1}}\mathbf{c}_3}{{{\mathbf{c}_3}^{\text{H}}}{{\mathbf{\Sigma }}_{\text{2}}}\mathbf{c}_3+\sigma _{2}^{2}}>{{\tau }_{3}}-{{\tau }_{1}},\,\,\,\,
     \frac{{{\mathbf{c}_3}^{\text{H}}}{{\mathbf{\Phi }}_{2}}\mathbf{c}_3}{{{\mathbf{c}_3}^{\text{H}}}{{\mathbf{\Sigma }}_{2}}\mathbf{c}_3+\sigma _{2}^{2}}-\frac{{{\mathbf{c}_3}^{\text{H}}}{{\mathbf{\Omega }}_{2}}\mathbf{c}_3}{{{\mathbf{c}_3}^{\text{H}}}{{\mathbf{\Sigma }}_{\text{1}}}\mathbf{c}_3+\sigma _{1}^{2}}>{{\tau }_{4}}-{{\tau }_{2}}\}.
    \end{aligned}
\end{equation}
Employing the SDR method by defining $\textbf{C}_3 = \textbf{c}_3 \textbf{c}_3^\text{H}$ and also using the Charnes-Cooper transformation by defining $\textbf{C}_3 = \frac{\textbf{Z}_3}{\zeta_3}$, for a known value of $t_3$, the problem (\ref{subp32}) can be written as,
\begin{equation}\label{subp3aftercooper}
    \begin{aligned}
   \underset{{{\mathbf{Z}}_{3}},\,{{\zeta }_{3}}}{\mathop{\text{max}}}\,\text{Tr}\left( {{\mathbf{\Phi }}_{31}}{{\mathbf{Z}}_{3}} \right)
     &\,\,\,\,\text{s}\text{.t.}\,\,\,\,\{\text{Tr}\left( {{\mathbf{\Omega }}_{{R}}}{{\mathbf{Z}}_{3}} \right)\le {{\zeta }_{3}}{{\text{P}}_{{R}}},
     \,\,\,\,\text{Tr}\left( \left( {{\mathbf{\Phi }}_{32}}-{{t}_{3}}{{\mathbf{\Sigma }}_{\text{2}}} \right){{\mathbf{Z}}_{3}} \right)={{\zeta }_{3}}{{t}_{3}}\sigma _{2}^{2} ,\,\,\,\,\text{Tr}\left( {{\mathbf{\Sigma }}_{\text{1}}}{{\mathbf{Z}}_{3}} \right)+{{\zeta }_{3}}\sigma _{1}^{2}=1,\\
     &\text{Tr}\left( {{\mathbf{\Phi }}_{1}}{{\mathbf{Z}}_{3}} \right)-\frac{\text{Tr}\left( {{\mathbf{\Omega }}_{1}}{{\mathbf{Z}}_{3}} \right)}{\text{Tr}\left( {{\mathbf{\Sigma }}_{\text{2}}}{{\mathbf{Z}}_{3}} \right)+{{\zeta }_{3}}\sigma _{2}^{2}}>{{\tau }_{3}}-{{\tau }_{1}},
     \,\,\,\,\frac{\text{Tr}\left( {{\mathbf{\Phi }}_{2}}{{\mathbf{Z}}_{3}} \right)}{\text{Tr}\left( {{\mathbf{\Sigma }}_{2}}{{\mathbf{Z}}_{3}} \right)+{{\zeta }_{3}}\sigma _{2}^{2}}-\text{Tr}\left( {{\mathbf{\Omega }}_{2}}{{\mathbf{Z}}_{3}} \right)>{{\tau }_{4}}-{{\tau }_{2}}, \\
     & {{\zeta }_{3}}>0,~~\mathbf{Z}_3\succeq0\}.
    \end{aligned}
\end{equation}
In order to convert (\ref{subp3aftercooper}) to the standard SDP form, defining $t_4= \text{Tr}\left( \mathbf{\Sigma} _2 \mathbf{Z}_3 \right)+\zeta_3\sigma _2^2$, the above problem can be solved for a set of values of $t_4$, where the one with maximum objective function is chosen as the best solution of (\ref{subp3aftercooper}). In fact, a two-dimensional search over values of $t_3$ and $t_4$ should be carried out to get the optimal solution of (\ref{subp3aftercooper}), where in the Appendix, using feasibility studies, upper bounds for $t_3$ and $t_4$ are provided to limit the search space. Mathematically speaking, the following problem should be solved for each value of $t_3$ and $t_4$,
\begin{equation}\label{subp3aftercoopert4}
\begin{aligned}
\underset{{{\mathbf{Z}}_{3}},\,{{\zeta }_{3}}}{\mathop{\text{max}}}\,\text{Tr}\left( {{\mathbf{\Phi }}_{31}}{{\mathbf{Z}}_{3}} \right)
 \,\,\,\,\text{s}\text{.t}\,\,\,\,&\{\text{Tr}\left( {{\mathbf{\Omega }}_{{R}}}{{\mathbf{Z}}_{3}} \right)\le {{\zeta }_{3}}{{\text{P}}_{{R}}},
 \,\,\,\,\text{Tr}\left( \left( {{\mathbf{\Phi }}_{32}}-{{t}_{3}}{{\mathbf{\Sigma }}_{\text{2}}} \right){{\mathbf{Z}}_{3}} \right)={{\zeta }_{3}}{{t}_{3}}\sigma _{2}^{2},
 \,\,\,\,\text{Tr}\left( {{\mathbf{\Sigma }}_{\text{1}}}{{\mathbf{Z}}_{3}} \right)+{{\zeta }_{3}}\sigma _{1}^{2}=1, \\
 &\text{Tr}\left( {{\mathbf{\Phi }}_{1}}{{\mathbf{Z}}_{3}} \right)-\frac{\text{Tr}\left( {{\mathbf{\Omega }}_{1}}{{\mathbf{Z}}_{3}} \right)}{{{t}_{4}}}>{{\tau }_{3}}-{{\tau }_{1}},
 \,\,\,\,\frac{\text{Tr}\left( {{\mathbf{\Phi }}_{2}}{{\mathbf{Z}}_{3}} \right)}{{{t}_{4}}}-\text{Tr}\left( {{\mathbf{\Omega }}_{2}}{{\mathbf{Z}}_{3}} \right)>{{\tau }_{4}}-{{\tau }_{2}},\\
 &\text{Tr}\left( {{\mathbf{\Sigma }}_{\text{2}}}{{\mathbf{Z}}_{3}} \right)+{{\zeta }_{3}}\sigma _{2}^{2}={{t}_{4}},\,\,\,\,{{\zeta }_{3}}>0,~~\mathbf{Z}_3\succeq0\}.
\end{aligned}
\end{equation}
\begin{remark}
The greatest value among the best solutions of (\ref{subp1aftercooper2}), (\ref{subp2aftercooper}), and (\ref{subp3aftercoopert4}) is a close-to-optimal solution of the main problem (\ref{main_problem}). According to simulation results, in most channel realizations, the optimal solution of the equivalent SDR problem maximizing the objective function in (\ref{main_problem}) is of rank one. Thus in most cases, the optimal solution of the main problem is achieved. Simulation results confirm this assertion, as the obtained solution closely follows the objective function of the equivalent SDR problem, where according to Remark 1, it is an upper bound of the main problem.
\end{remark}
\begin {remark}Complexity analysis: For an SDR problem whose unknown matrix is of $q\times q$ dimension and it has $p$ linear constraints, the worst case complexity is $\mathcal{O}\left( \max {{\left\{ p,q \right\}}^{4}}{{q}^{\frac{1}{2}}}\log \left( 1/\varepsilon  \right) \right)$, where $\varepsilon > 0$ is the solution accuracy \cite{sroqop}. Due to the employing the SDR technique in this paper, the complexity of three presented sub-problems, the main problem, and the investigated problem in \cite{base} when using an N-antenna relay can be calculated as table \ref{table:1}. In this table, $M_1$ is the number of searches over $t_1$ through $t_4$, while $M_2$ is the number of steps to get the best solution according to the bisection algorithm presented in \cite{base}. Also, we have ${{J}_{1}}=\max {{\left\{ 5,\left( {{N}^{2}}-2 \right) \right\}}^{4}}{{\left( {{N}^{2}}-2 \right)}^{\frac{1}{2}}}\log \left( 1/\varepsilon  \right)$ and ${{J}_{2}}=\max {{\left\{ 7,\left( {{N}^{2}}-2 \right) \right\}}^{4}}{{\left( {{N}^{2}}-2 \right)}^{\frac{1}{2}}}\log \left( 1/\varepsilon  \right)$. It is worth mentioning that for large values of $N$, the proposed approach scales as $\mathcal{O}\left( M_1^2N^9 \right)$, while the method in \cite{base} scales as $\mathcal{O}\left( M_2N^9 \right)$ where for the same number of search steps, i.e., $M=M_1=M_2$, the complexity of the proposed approach is $M$ times greater than that of \cite{base}.
\end{remark}
\begin{table}
\centering
\begin{tabular}{ |c|c| }
\hline
 Problem & Complexity  \\
\hline
 Sub-problem1 & $\mathcal{O}\left( {{M}_{1}}\max {{\left\{ 5,\left( {{N}^{2}}-2 \right) \right\}}^{4}}{{\left( {{N}^{2}}-2 \right)}^{\frac{1}{2}}}\log \left( 1/\varepsilon  \right) \right)$  \\
\hline
  Sub-problem2 & $\mathcal{O}\left( {{M}_{1}}\max {{\left\{ 5,\left( {{N}^{2}}-2 \right) \right\}}^{4}}{{\left( {{N}^{2}}-2 \right)}^{\frac{1}{2}}}\log \left( 1/\varepsilon  \right) \right)$  \\
  \hline
 Sub-problem3 & $\mathcal{O}\left( M_{1}^{2}\max {{\left\{ 7,\left( {{N}^{2}}-2 \right) \right\}}^{4}}{{\left( {{N}^{2}}-2 \right)}^{\frac{1}{2}}}\log \left( 1/\varepsilon  \right) \right)$  \\
   \hline
 Main problem & $\mathcal{O}\left( \max \left\{ {{M}_{1}}{{J}_{1}},{{M}_{1}}{{J}_{1}},M_{1}^{2}{{J}_{2}} \right\} \right)= \mathcal{O}\left(M_{1}^{2}{{J}_{2}} \right)$  \\
   \hline
 Problem in \cite{base}  & $\mathcal{O}\left( {{M}_{2}}\max {{\left\{ 3,\left( N^2-2 \right) \right\}}^{4}}{{\left( N^2-2 \right)}^{\frac{1}{2}}}\log \left( 1/\varepsilon  \right) \right)$  \\
 \hline
\end{tabular}
\caption{The complexity analysis}\label{table:1}
\end{table} 
\section{NUMERICAL RESULTS}\label{sec:simul}
\begin{figure}[t]
\centering
 \includegraphics[scale = .5]{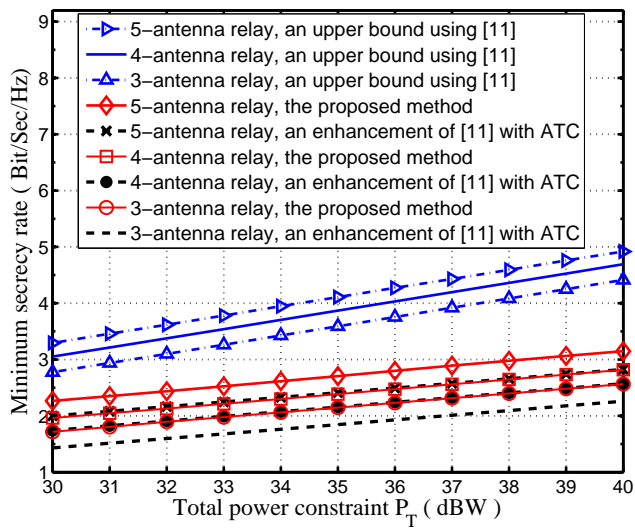}
  \caption{The average value of the minimum achievable secrecy rate vs. the total power constraint in a network with a 3-, 4-, and 5-antenna relay.}\label{All}
\end{figure}

\begin{figure}[t]
\centering
 \includegraphics[scale = .5]{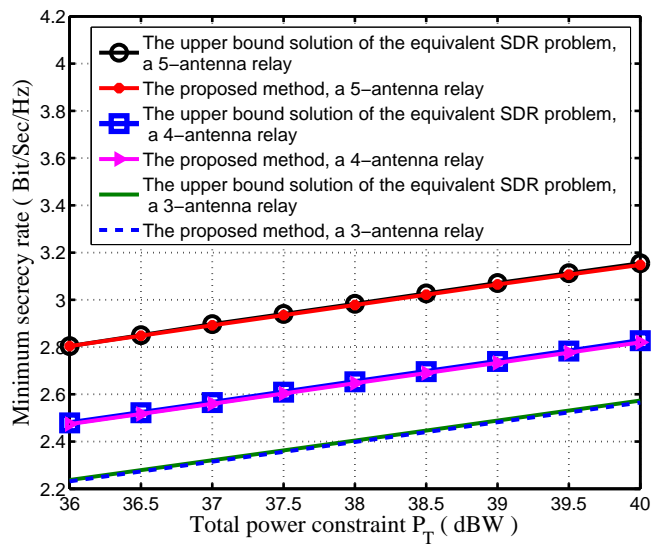}
  \caption{ The upper bound solutions of the equivalent SDR problem and the proposed method vs. the total power constraint in a network using a 3-, 4-, and 5-antenna relay.}\label{upperlower}
\end{figure}

\begin{figure}[t]
\centering
 \includegraphics[scale = .4]{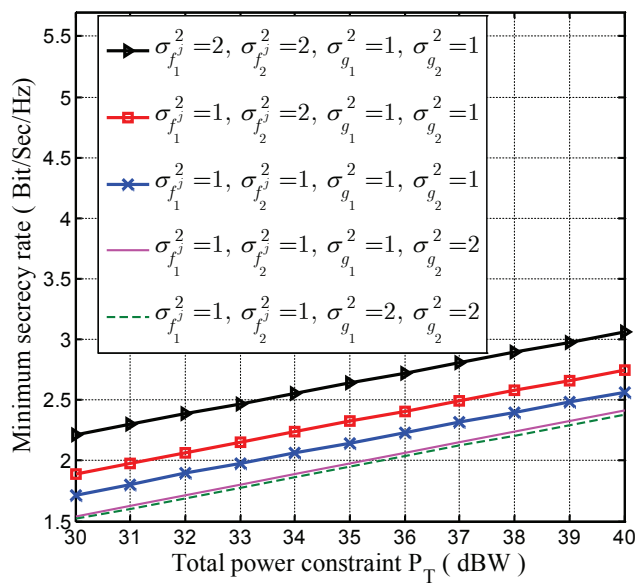}
  \caption{The average value of the minimum achievable secrecy rate obtained by the proposed approach vs. the total power constraint in a network with a 3-antenna relay for different settings of $\sigma_{f_1^{ j}}^{2}$, $\sigma_{f_2^{ j}}^{2}$, $\sigma_{g_1}^{2}$, and $\sigma_{g_2}^{2}$.}\label{difsettingsigma}
\end{figure}

\begin{figure}[t]
\centering
 \includegraphics[scale = .5]{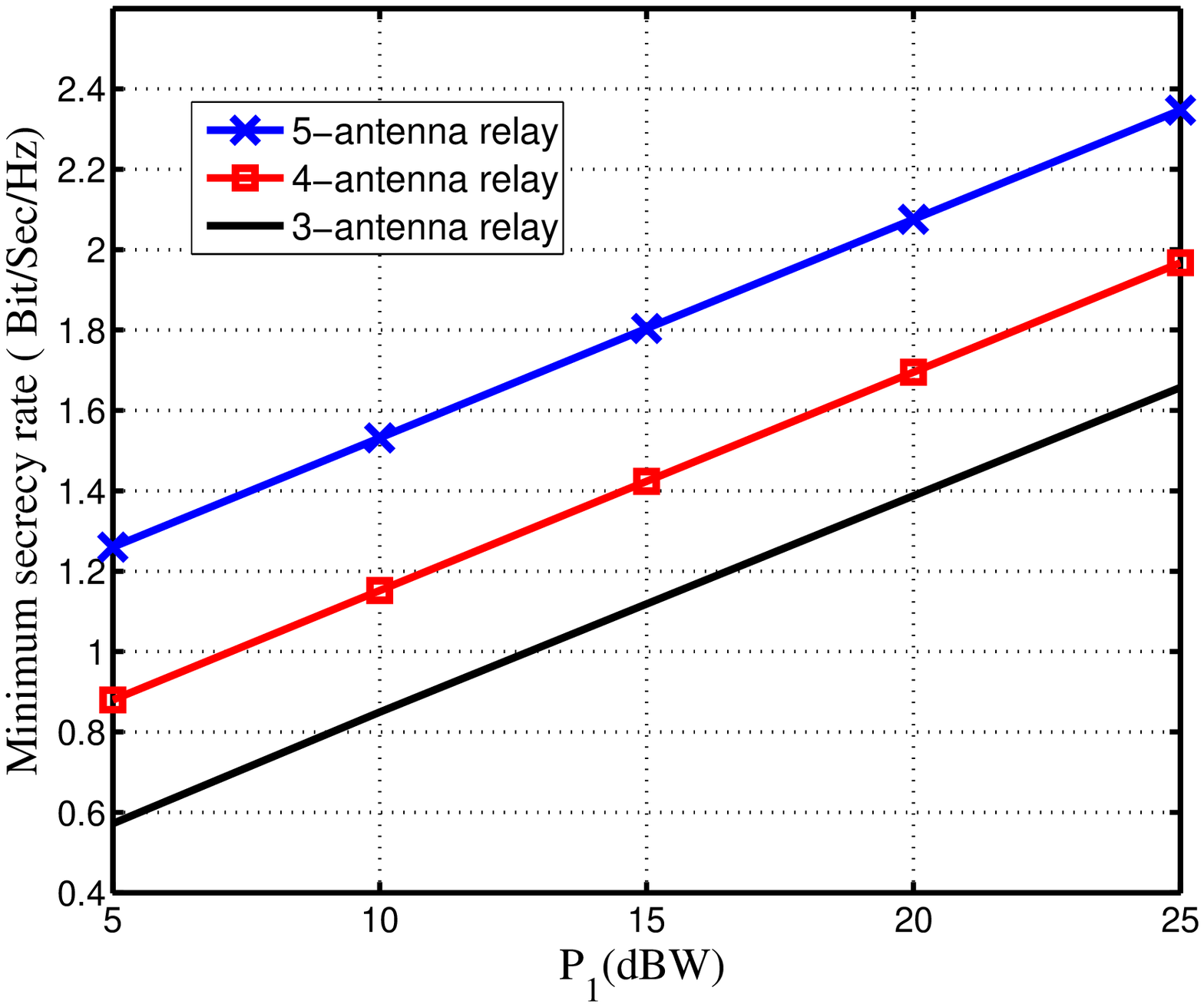}
  \caption{The average value of the minimum achievable secrecy rate vs. ${{\text{P}}_{{1}}}$ in a network using a 3-, 4-, and 5-antenna relay with ${{\text{P}}_{{R}}}=20$ dBW and ${{\text{P}}_{{2}}}=15$ dBW.}\label{Relay_cons}
\end{figure}

This section aims to compare the minimum achievable secrecy rate of the proposed method with the method given in \cite{base}, which to the best of authors' knowledge is the best-known method addressed in the literature. This gives an indication regarding the advantage of the current study. Throughout the simulations, all channel coefficients are assumed to be i.i.d. complex zero-mean Gaussian random variables. The variances of the $f_1^{ j}$, $f_2^{ j}$, ${g_1}$, and ${g_2}$, where $j=1,2,...,N$, are indicated by $\sigma_{f_1^{ j}}^{2}$, $\sigma_{f_2^{ j}}^{2}$, $\sigma_{g_1}^{2}$, and $\sigma_{g_2}^{2}$, which in Figs. \ref{All}, \ref{upperlower}, and \ref{Relay_cons} are considered to be of unit value, however, they have different values in Fig. \ref{difsettingsigma}. Moreover, all noise powers, including the noise power at transmitting ends ($\sigma _{k}^{2}$ with node number $k\in \{1,2\}$), the relay ($\sigma _{R}^{2}$), and the eavesdropper ($\sigma _{E,k'}^{2}$ with time index $k' \in \{1,2\}$), are assumed to be of unit power.

Similar to what is done in~\cite{base}, to have a fair comparison result, for the results presented in Figs. \ref{All}, \ref{upperlower}, and \ref{difsettingsigma}, a quarter of total power budget $P_T$ is allocated to both source 1 ($P_1$) and source 2 ($P_2$), while the remaining power is allocated to the relay ($\text{P}_{R}$), thus we have ${{\text{P}}_{1}}={{\text{P}}_{2}}=\frac{{{\text{P}}_{\text{T}}}}{4}$ and ${{\text{P}}_{R}}=\frac{{{\text{P}}_{\text{T}}}}{2}$. It is worth mentioning that for each value of $P_T$, the corresponding optimization problem is solved for 400 channel realizations and the average solution is represented in numerical results. In our proposed approach, the solution of the optimization problem is the greatest value among the solutions of three sub-problems. Furthermore, the step size values for search over $t_1$ to $t_4$ are set as one five-hundredth ($\frac{1}{500}$) of the associated interval.

It should be noted that~\cite{base} considers some single-antenna relays, while the current study assumes a multi-antenna relay. Thus, for the sake of fair comparison, the work done in \cite{base} is extended to a multi-antenna single relay network. Moreover, as is mentioned in Section~\ref{sec:intro}, \cite{base} merely considers two constraints $R_1\leq R_1^{up}$ and $R_2\leq R_2^{up}$, while the third constraint $R_1+R_2\leq R_{sum}^{up}$ is not included in the constraints when attempting to maximize $\min(R_1,R_2)$, thus the objective function in \cite{base} would be an upper bound since the incorporated feasible set, i.e., the feasible set identified with the first two constraints, subsumes the feasible set $\mathcal{R}$ defined in (\ref{ASR Region}) with three mentioned constraints.

Nevertheless, one can readily compute $\bold{W}$ by \cite{base} and then, using that, identify the secrecy rate region given in~(\ref{ASR Region}), where the minimum ASR is calculated over this region through finding the intersection of line $R_1=R_2$ with the borders of this region. Accordingly, the result can be treated as an inner bound. In the obtained results, this method is called the enhancement of~\cite{base} by applying the third constraint (ATC).

Fig. \ref{All} illustrates the upper bound curves based on the method discussed in \cite{base}, the proposed method, and an enhancement of \cite{base} with ATC for a network exploiting a 3-, 4-, and 5-antenna relay. Accordingly, it can be observed that the performance of the proposed method using m-antenna is approximately equivalent to that of the enhancement of \cite{base} with ATC when $m+1$ antennas are employed at the relay. Also, as is inferred from Fig. \ref{All}, the proposed approach has superior performance w.r.t. \cite{base}. More specifically, with the same achievable secrecy rate, the consumed power of the proposed approach is about 3 dBW less than that of the enhancement of \cite{base} with ATC, but at the expense of higher computational complexity. In addition, the associated curves of the proposed method give an indication regarding the impact of increasing the number of antennas at the relay, showing that for the same minimum achievable secrecy rate, there is a dramatic decrease in the consumed power.

Fig. \ref{upperlower} is provided to illustrate that the proposed approach is a close-to-optimal solution to confirm the assertion of Remark 2 in Section\ref{sec:prob}. Accordingly, the solution of the corresponding SDR problem, which is actually an upper bound, is compared to that of the proposed approach, showing that they closely follow each other.

Fig. \ref{difsettingsigma} illustrates the minimum ASR of the proposed method for various values of $\sigma_{f_\text{i}^{ j}}^{2}$ and $\sigma_{g_\text{i}}^{2}$ for $\text{i}=1,2$ in a network utilizing a 3-antenna relay. Since $\sigma_{f_\text{i}^{ j}}^{2}$ and $\sigma_{g_\text{i}}^{2}$ can be, respectively, thought as a measure of closeness of the $j^{th}$ relay's antenna and the eavesdropper to the source $S_\text{i}$, changing these values gives an indication regarding the impact of node distances on the resulting minimum ASR. Referring to Fig. \ref{difsettingsigma}, considering the case of $\sigma_{f_\text{1}^{ j}}^{2}=\sigma_{f_\text{2}^{ j}}^{2}=\sigma_{g_\text{1}}^{2}=\sigma_{g_\text{2}}^{2}=1$ as the benchmark, as is expected, one can readily observe that increasing both values of $\sigma_{f_\text{1}^{ j}}^{2}$ and $\sigma_{f_\text{2}^{ j}}^{2}$ increases the minimum ASR. Also, increasing both channel strengths from users to the eavesdropper, i.e., $\sigma_{g_\text{1}}^{2}$ and $\sigma_{g_\text{2}}^{2}$, has a negative impact on the resulting minimum ASR. However, increasing the channel strength of one user to relay does not greatly influence the result. On the other hand, increasing the channel strength from one user to the eavesdropper has approximately the same negative impact w.r.t. the case of increasing both user-to-eavesdropper channel strengths. This is due to the fact that the objective is maximizing the minimum ASR and in this case the user with the worst-case channel strength to the eavesdropper acts as a bottleneck.

Fig. \ref{Relay_cons} depicts the impact of increasing transmit power of $S_1$ on the minimum ASR when the available power at the relay and the transmit power of $S_1$ are, respectively, set to ${{\text{P}}_{{R}}}=20$ dBW and ${{\text{P}}_{{2}}}=15$ dBW, showing that the minimum ASR improves with increasing ${{\text{P}}_{{1}}}$. It should be noted that as the signal vectors received at the relay are mostly from different directions, the AF relay can simply adjust its transmit beamforming matrix in such a way that the user's signal vector with greater power does not consume most of relay's power. Thus, increasing power does not have a negative impact. This confirms the assumption that full power transmission gives the best result.
\section{CONCLUSION}
This paper aimed at maximizing the minimum achievable secrecy rate of a network assisted by  a two-way multi-antenna relay, assuming transmission occurs in two hops. Throughout the first hop, the multiple access phase, the relay as well as eavesdropper receive combined messages from both transmitters, while in the second hop, the broadcast phase, the relay makes use of null space beamforming to transmit the combined information in the null space of the eavesdropper's channel to enhance the secrecy rate. Accordingly, a proper beamforming strategy at the relay with the view of increasing the minimum achievable secrecy rate is tackled and the result is compared to that of \cite{base}, which is the best-known method in the literature, showing that the proposed method outperforms existing work. Moreover, it is numerically shown that the proposed approach closely follows the upper bound justifying the superior performance of the proposed method. 
\section{APPENDIX}
This section aims to provide upper bounds for parameters $t_1$ to $t_4$. To this end, referring to (\ref{subp1aftercooper2}), the optimum value of the following optimization problem can be considered as the upper bound of $t_1$.
\begin{equation}\label{uppert1}
\begin{aligned}
\underset{{{\mathbf{Z}}_{1}},\,{{\text{ }\!\!\zeta\!\!\text{ }}_{1}}}{\mathop{\max }}~\text{Tr}\left( {{\mathbf{\Sigma }}_{\text{2}}}{{\mathbf{Z}}_{1}} \right)+{{\text{ }\!\!\zeta\!\!\text{ }}_{1}}\sigma _{2}^{2}
 \,\,\,\text{s}\text{.t}\text{.}\,\,\,\{\text{Tr}\left( {{\mathbf{\Omega }}_{\text{R}}}{{\mathbf{Z}}_{1}} \right)\le {{\text{ }\!\!\zeta\!\!\text{ }}_{1}}{{\text{P}}_{\text{R}}},
 \,\,\,\text{Tr}\left( {{\mathbf{\Sigma }}_{\text{1}}}{{\mathbf{Z}}_{1}} \right)+{{\zeta }_{1}}\sigma _{1}^{2}=1,
 \,\,\,{{\text{ }\!\!\zeta\!\!\text{ }}_{1}}>0\}.
\end{aligned}
\end{equation}
Noting (\ref{subp1aftercooper2}) and (\ref{subp3aftercoopert4}), the optimum solution of (\ref{uppert1}) can provide an upper bound for both $t_1$ and $t_4$. Also, referring to (\ref{subp2aftercooper}), the optimum value of the following optimization problem can be considered as the upper bound of $t_2$.
\begin{equation}\label{uppert2}
\begin{aligned}
\underset{{{\mathbf{Z}}_{2}},{{\zeta }_{2}}}{\mathop{\max }}~\text{Tr}\left( {{\mathbf{\Sigma }}_{1}}{{\mathbf{Z}}_{2}} \right)+{{\zeta }_{2}}\sigma _{1}^{2}
\,\,\,\text{s}\text{.t}\text{.}\,\,\,\{\text{Tr}\left( {{\mathbf{\Omega }}_{\text{R}}}{{\mathbf{Z}}_{2}} \right)\le {{\zeta }_{2}}{\text{P}_{\text{R}}},
 \,\,\,\text{Tr}\left( {{\mathbf{\Sigma }}_{2}}{{\mathbf{Z}}_{2}} \right)+{{\zeta }_{2}}\sigma _{2}^{2}=1,\,\,\,{{\zeta }_{2}}>0\}.
\end{aligned}
\end{equation}
To find the upper bound of $t_3$ in (\ref{subp32}), the following optimization problem is considered,
\begin{equation}\label{uppert31}
 \begin{aligned}
\underset{\mathbf{c}}{\mathop{\max }}\,\,\,\,\frac{\mathbf{c}_{\,}^{\,\text{H}}{{\mathbf{\Phi }}_{32}}\mathbf{c}}{\mathbf{c}_{{}}^{\,\text{H}}{{\mathbf{\Sigma }}_{\text{2}}}\mathbf{c}+\sigma _{2}^{2}}
\,\,\,\text{s}\text{.t}\text{.}\,\,\,\{\mathbf{c}_{\,}^{\,\text{H}}{{\mathbf{\Omega }}_{\text{R}}}\mathbf{c}\le {{\text{P}}_{\text{R}}}\},
 \end{aligned}
\end{equation}
where after employing the SDR technique and using the Charnes-Cooper method, the upper bound of $t_3$ is numerically derived. 


\begin{thebibliography}{1}
\bibitem{Wyner}
Wyner, A.D.: 'The wiretap channel', \emph{Bell Syst. Tech. J.}, 1975, \textbf{54}, (8), pp. 1355--1387
\bibitem{ucdpisd}
Sendonaris, A., Erkip, E., Aazhang, B.: 'User cooperation diversity-part I: system description', \emph{IEEE Trans. Commun.}, 2003, \textbf{51}, (11), pp. 1927-–1938
\bibitem{khajenouri}
Khajehnouri, N., Sayed, A. H.: 'Distributed MMSE relay strategies for wireless sensor networks', \emph{IEEE Trans. Signal Process.}, 2007, \textbf{55}, (7), pp. 3336–-3348
\bibitem{treccfs}
Lai, L., Gamal, H. E.: 'The relay eavesdropper channel: Cooperation for secrecy', \emph{IEEE Trans. Inf. Theory}, 2008, \textbf{54}, (9), pp. 4005–-4019
\bibitem{orsfplsicwn}
Zou, Y., Wang, X., Shen, W.: 'Optimal relay selection for physical-layer security in cooperative wireless
networks', \emph{IEEE J. Sel. Areas Commun.}, 2013, \textbf{31}, (10), pp. 2099–-2111
\bibitem{soodharswrswec}
Jindal, A., Kundu, C., Bose, R.: 'Secrecy outage of dual-hop AF relay system with relay selection
without eavesdropper’s CSI ', \emph{IEEE Commun. Lett.}, 2014, \textbf{18}, (10), pp. 1759–-1762
\bibitem{spomamrnwors}
Wang, L., Xu, S., Yang, W., Yang, W., Cai, Y.: 'Security performance of multiple antennas
multiple relaying networks with outdated relay selection', Proc. IEEE Wireless Commun. Signal Process. (WCSP), Hefei, Anhui, China, October 2014, pp. 1–-6
\bibitem{sepfhfrc}
Rankov, B., Wittneben, A.: 'Spectral efficient protocols for half-duplex fading relay channels
networks', \emph{IEEE J. Sel. Areas Commun.}, 2007, \textbf{25}, (2), pp. 379–-389
\bibitem{wncbaaffbdtf}
Popovski, P., Yomo, H.: 'Wireless network coding by amplify-and-forward for bi-directional traffic flows', \emph{IEEE Commun. Lett.}, 2007, \textbf{11}, (1), pp. 16–-18
\bibitem{dbfplsotwrn}
Wang, H.-M., Yin, Q., Xia, X.-G.: 'Distributed beamforming for physical-layer security of two-way relay networks', \emph{IEEE Trans. Signal Process.}, 2012, \textbf{60}, (7), pp. 3532–-3545
\bibitem{base}
Yang, Y., Sun, C., Zhao, H., Long, H., Wang, W.: 'Algorithms for secrecy guarantee with null space
beamforming in two-way relay networks', \emph{IEEE Trans. Signal Process.}, 2014, \textbf{62}, (8), pp. 2111–-2126
\bibitem{hcbajfplsotwrn}
Wang, H.-M., Luo, M., Yin, Q., Xia, X.-G.: 'Hybrid cooperative beamforming and jamming for physical-layer
security of two-way relay networks', \emph{IEEE Trans. Inf. Forensics Sec.}, 2013, \textbf{8}, (12), pp. 2007–-2020
\bibitem{rcbaandfplsiamamrn}
Li, Q., Yang, Y., Ma, W.-K., Lin, M., Ge, J., Lin, J.: 'Robust cooperative beamforming and artificial noise design for physical-
layer secrecy in AF multi-antenna multi-relay networks', \emph{IEEE Trans. Signal Process.}, 2015, \textbf{63}, (1), pp. 206–-220
\bibitem{jsrpapafsaafmrn}
Wang, H.-M., Liu, F., Xia, X.-G.: 'Joint source-relay precoding and power allocation for secure amplify-and-forward MIMO
relay networks', \emph{IEEE Trans. Inf. Forensics Sec.}, 2014, \textbf{9}, (8), pp. 1240–-1250
\bibitem{cjfscimrn}
Huang, J., Swindlehurst, A. L.: 'Cooperative jamming for secure communications in MIMO relay networks', \emph{IEEE Trans. Signal Process.}, 2011, \textbf{59}, (10), pp. 4871–-4884
\bibitem{Aylin}
Tekin, E., Yener, A.: 'The general Gaussian multiple-access and two-way wire tap channels: achievable rates and
cooperative jamming', \emph{IEEE Trans. Inf. Theory}, 2008, \textbf{54}, (6), pp. 2735–-2751
\bibitem{cvx}
Grant, M., Boyd, S.: 'CVX: MATLAB software for disciplined convex programming', September 2010. Available at http://cvxr.com/cvx
\bibitem{Charnescooper}
Charnes, A., Cooper, W.W.: 'Programming with linear fractional functionals', \emph{Nav. Res. Logist. Q.}, 1962, \textbf{9}, (3-4), pp. 181-–186
\bibitem{sroqop}
Luo, Z.-Q., Ma, W.-K., So, A. M.-C., Ye, Y., Zhang, S.: 'Semidefinite relaxation of quadratic optimization problems', \emph{IEEE Signal Process. Mag.}, 2010, \textbf{27}, (3), pp. 20–-34 
\end{thebibliography}
\end{document}